\documentclass[a4paper,twocolumn,11pt,accepted=2025-04-01]{quantumarticle}
\pdfoutput=1

\usepackage[usenames, dvipsnames]{color}
\usepackage{tikz}
\usetikzlibrary{shapes,snakes,backgrounds,fit,decorations.pathreplacing}
\usepackage{bbm, bm, amsbsy}
\usepackage{amsthm}
\usepackage{amssymb}
\usepackage{amsmath}
\usepackage{dsfont} 
\usepackage{graphicx} 
\usepackage{epsfig}
\usepackage{epstopdf}

\usepackage[numbers,sort&compress]{natbib}
\usepackage[english]{babel}
\usepackage[T1]{fontenc}

\usepackage{soul}

\usepackage{hyperref}
\hypersetup{
	colorlinks,
	linkcolor={blue},
	citecolor={blue},
	urlcolor={blue}
}
\usepackage{ragged2e}
\usepackage[capitalise]{cleveref}
\usepackage[normalem]{ulem}
\usepackage{units}
\usepackage{braket}
\usepackage{float}

\usepackage{mathrsfs}
\usepackage{braket}
\usepackage{booktabs}
\usepackage{tikz}
\usepackage{lipsum}

\usepackage[explicit]{titlesec}
\usepackage{textcase}

\newcommand{\subtiny}[3]{\ensuremath{_{\hspace{#1 pt}\protect\raisebox{#2 pt}{\tiny{$ #3$}}}}}
\newcommand{\suptiny}[3]{\ensuremath{^{\hspace{#1 pt}\protect\raisebox{#2 pt}{\tiny{$ #3$}}}}}
\newcommand{\tr}{\textnormal{Tr}}
\newcommand{\scpr}[2]{\ensuremath{\left\langle\right.\hspace*{-1pt} #1 \hspace*{-1pt}\left|\right.\hspace*{-1pt} #2 \hspace*{-1pt}\left.\right\rangle}}

\begin{document}

\title{Multi-copy activation of genuine multipartite entanglement in continuous-variable systems}

\author{Kl{\'a}ra Baksov{\'a}}
\email{klara.baksova@tuwien.ac.at}
\affiliation{Atominstitut, Technische Universit{\"a}t Wien, Stadionallee 2, 1020 Vienna, Austria}
\orcid{0009-0009-8944-6044}
\thanks{K. B. and O. L. contributed equally.}
\author{Olga Leskovjanov{\'a}}
\email{ola.leskovjanova@gmail.com}
\orcid{0000-0002-3638-3627}
\affiliation{Department of Optics, Palack\'y University, 17. listopadu 12, 771~46 Olomouc, Czech Republic}

%
\author{Ladislav Mi\v{s}ta, Jr.}
\email{mista@optics.upol.cz}
\orcid{0000-0002-1979-7617}
\affiliation{Department of Optics, Palack\'y University, 17. listopadu 12, 771~46 Olomouc, Czech Republic}
\author{Elizabeth Agudelo}
\email{elizabeth.agudelo@tuwien.ac.at}
\orcid{0000-0002-5604-9407}
\affiliation{Atominstitut, Technische Universit{\"a}t Wien, Stadionallee 2, 1020 Vienna, Austria}
\author{Nicolai Friis}
\email{nicolai.friis@tuwien.ac.at}
\orcid{0000-0003-1950-8640}
\affiliation{Atominstitut, Technische Universit{\"a}t Wien, Stadionallee 2, 1020 Vienna, Austria}
\maketitle

\begin{abstract}
Multi-copy activation of genuine multipartite entanglement (GME) is a phenomenon whereby multiple copies of biseparable but fully inseparable states can exhibit GME. This was shown to be generically possible in finite dimensions. Here, we extend this analysis to infinite dimensions. We provide examples of GME-activatable non-Gaussian states. For Gaussian states, we apply a necessary biseparability criterion for the covariance matrix and show that it cannot detect GME activation. We further identify fully inseparable Gaussian states that satisfy the criterion but show that multiple and, in some cases, even single copies are GME. Thus, we show that the covariance-matrix biseparability criterion is not sufficient even for Gaussian states.
\end{abstract}

\section{Introduction}
Entanglement stands as a key phenomenon in quantum physics, playing an essential role in the advancement of contemporary quantum technologies. Initially, attention was largely centered on two-party cases, but multipartite entanglement has become highly significant, both practically and fundamentally~\cite{GreenbergerHorneShimonyZeilinger,PanBouwmeesterDaniellWeinfurterZeilinger2000}. In experiments distributing quantum states among various parties, often multiple identical copies of these states are shared. Understanding entanglement properties in multi-copy situations is, therefore, essential.

\newpage
A~known feature of the two-party case is that bipartite separability is tensor stable: bipartite entanglement cannot be established between two parties by sharing multiple copies of separable states. This trivially extends to \emph{partition-separable} states of more than two parties, i.e., states separable with respect to a fixed partitioning of the parties into two groups. However, the same is not true for more complex states of multiple parties. States that are mixtures of partition-separable states for different partitions are called \emph{biseparable}, and they form a convex set. All quantum states that lie outside of this convex set are called \textit{genuinely multipartite entangled} (GME)~\cite{DuerVidalCirac2000}. 
\textit{Multi-copy activation of GME} is an initially perhaps counterintuitive phenomenon that occurs for states that are biseparable but not partition-separable:
Even though a single copy of a state might be biseparable, several identical copies of such a state can feature GME with respect to the parties sharing these copies. 

First remarked upon in Ref.~\cite{HuberPlesch2011}, for two copies of a specific four-qubit state, GME activation was investigated more comprehensively in Ref.~\cite{YamasakiMorelliMiethlingerBavarescoFriisHuber2022}. There, upper bounds were provided for the number of copies maximally needed to activate GME for a family of $N$-qubit states, and it was shown that GME activation could even occur for biseparable states with positive partial transpose across all cuts, i.e., states with no distillable entanglement. These results were generalized in Ref.~\cite{PalazuelosDeVicente2022} for all finite-dimensional multipartite states, showing that all \textit{biseparable} but \textit{not partition separable} states are GME-activatable, but an unbounded number of copies may be required. 

Here, we investigate GME activation in infinite dimensions, specifically in continuous-variable (CV) systems~\cite{vanLoockFurusawa,BraunsteinVanLoock2005}, and we focus on two categories: non-Gaussian~\cite{Walschaers2021} and Gaussian states~\cite{WeedbrookPirandolaGarciaPatronCerfRalphShapiroLloyd2012,AdessoRagyLee2014}. For non-Gaussian states, there are multipartite states that have non-zero overlap with only finitely many Fock states. These states, as well as all their marginals, can be fully represented on finite-dimensional Hilbert spaces, and the results for GME activatability from Ref.~\cite{PalazuelosDeVicente2022} apply directly. However, not all non-Gaussian states are of this form. 

As a first main result, we show that GME activation occurs also in infinite-dimensional systems. To do so, we extend arguments from Refs.~\cite{SperlingVogel2009,PalazuelosDeVicente2022} to show that all biseparable but not partition-separable states in infinite dimensions are GME activatable.

We then provide a specific example of GME activation for a family of biseparable non-Gaussian states with non-zero overlap with infinitely many Fock states. These are convex combinations of product states of two-mode squeezed vacua with a Fock state and are thus biseparable. To detect GME, we use the $k$-separability criterion from Ref.~\cite{GabrielHiesmayrHuber2010}, which reveals that two copies of the considered state are GME for a continuous range of squeezing parameters. 
 
The third and main focus of this work is GME activation for Gaussian states. Here, the challenge lies in determining if a given state is biseparable but not partition separable. Since Gaussian states are fully described by their first and second moments, and because the former can be freely adjusted by local unitaries (displacements), entanglement properties of Gaussian states are fully captured by their second moments, organized into a covariance matrix (CM). A Gaussian state with CM $\gamma$ is fully separable with respect to a partition into $N$ parties if and only if there exist CMs $\gamma\suptiny{0}{0}{(1)}$, $\cdots$, $\gamma\suptiny{0}{0}{(N)}$ corresponding to the $N$ subsystems satisfying $\gamma\geq\gamma\suptiny{0}{0}{(1)}\oplus\cdots\oplus\gamma\suptiny{0}{0}{(N)}$~\cite{WernerWolf2001, AndersPhDThesis2003}. 
A generalization can be found in Ref.~\cite{HyllusEisert2006}: for all biseparable states (BS) with CM $\gamma\subtiny{0}{0}{\mathrm{BS}}$ there exist CMs $\gamma\subtiny{0}{0}{M(i)}$ that are block-diagonal with respect to the partition $M(i)$ and probabilities $p_i$ with $\sum_i p_i=1$ and $0\leq p_i \leq 1$ such that $\gamma\subtiny{0}{0}{\mathrm{BS}}-\sum_i p_i\gamma\subtiny{0}{0}{M(i)}\geq0$.

In the context of this inequality, which we dub the \emph{CM biseparability criterion}, we present three main results: First, we show that the CM biseparability criterion is insufficient for detecting the potential activation of GME for any number of copies. That is, we prove that if the CM of the single copy satisfies the criterion, then so do the CMs of any number of identical copies. If, like its counterparts for bipartite or full separability, the CM biseparability criterion was indeed necessary and sufficient for the biseparability of Gaussian states, our first result would imply that GME activation is impossible for Gaussian states. 
However, as a second main result, we show that there exist Gaussian states that satisfy the CM biseparability criterion but which are, in fact, GME. As a corollary, we then show that this leads to the perhaps surprising conclusion that there exist Gaussian states that are GME even though the first and second moments that fully define them exactly match those of a biseparable but non-Gaussian state. 
To present these results, we first continue with brief expositions on multipartite entanglement and CV systems before returning to the CM biseparability criterion, along with the proofs and discussion of our main results. 

\section{Framework}

\subsection{Bipartite entanglement}

For two quantum systems with Hilbert spaces $\mathscr{H}_A$ and $\mathscr{H}_B$, respectively, a global pure state $\ket{\Psi}\subtiny{0}{0}{AB}\in\mathscr{H}_A\otimes\mathscr{H}_B$ is called \textit{separable} if and only if it can be written as a tensor product $\ket{\Psi}\subtiny{0}{0}{AB}=\ket{\psi}_A \otimes\ket{\phi}_B$. For mixed states represented by density operators $\rho=\sum_j p_j \ket{\varphi_j}\!\!\bra{\varphi_j}$, where the $p_j$ are probability weights fulfilling $\sum_j p_j=1$ and $\ket{\varphi_j}\in\mathscr{H}_A\otimes\mathscr{H}_B$, a global state $\rho\subtiny{0}{0}{AB}$ is separable if and only if it can be written as a convex combination of tensor products of density operators of the two subsystems, 
\begin{align}
    \rho\subtiny{0}{0}{AB} &=\sum_i \,p_i \,\rho_i\suptiny{0}{0}{A} \otimes \rho_i\suptiny{0}{0}{B}.
\end{align}
States that are not separable are called \textit{entangled}. 

\subsection{Mutlipartite entanglement}

In multipartite scenarios with $N$ parties and a Hilbert space $\mathscr{H}_N=\bigotimes_{i=1}^N\mathscr{H}_i$ one may investigate separability with respect to different \textit{partitions} $M$ as a grouping of the set $[N]:=\{1,\cdots, N\}$. Specifically, dividing $[N]$ into two or more disjoint subsets $m(i)\subset [N]$ whose union is $[N]$, i.e., $\bigcup_{i} m(i)=[N]$ and $m(i)$$\cap m(j)$$=\emptyset\,\forall\,i\neq j$, gives partition $M(i)=\{m(i)\}$. We then use the following terminology: A partition into $k$ subsets is called a $k$-partition, and a pure state in $\mathscr{H}_N$ is called \emph{$k$-separable} if it can be written as a tensor product of $k$ pure states for at least one $k$-partition $M(i)$, i.e.,
\begin{align}
    \ket{\Phi^{(k)}}=\bigotimes_{i=1}^k \ket{\phi_{m(i)}},
    \quad\ket{\phi_{m(i)}}\in\mathscr{H}_{m(i),}
\end{align}
where $\mathscr{H}_{m(i)}$ is the Hilbert space of subsystems from the set
$m(i)$.\\
A mixed state is called $k$-separable if it can be decomposed as a convex mixture of pure states that are (at least) $k$-separable, i.e.,
\begin{align}
    \rho^{(k)}=\sum_i p_i \ket{\Phi_i^{(k)}}\bra{\Phi_i^{(k)}}.
\end{align}
Note that the different terms of the decomposition may be $k$-separable with respect to different $k$-partitions.
Further, a \textit{k}-separable state of $N$ parties is called \textit{fully separable} for $k=N$, and it is called \emph{biseparable} for $k=2$. Any state that is separable with respect to any fixed partition is called \textit{partition separable}, whereas a state that is not separable with respect to any fixed partition is called \emph{fully inseparable}. 

 \begin{figure}[t]
        \centering
            \flushleft{(a)}\\
            \centering\includegraphics[width=0.75\columnwidth]{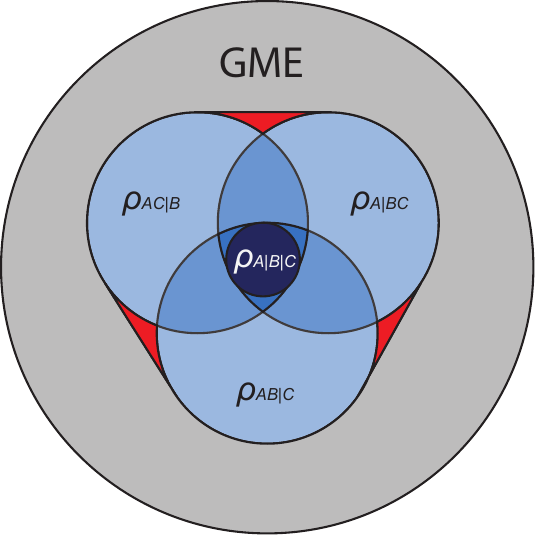}
            \flushleft{(b)}\hfill\\[-6mm]
            \centering\includegraphics[width=0.85\columnwidth,trim={4.5cm 2.2cm 4.5cm 2.3cm},clip]{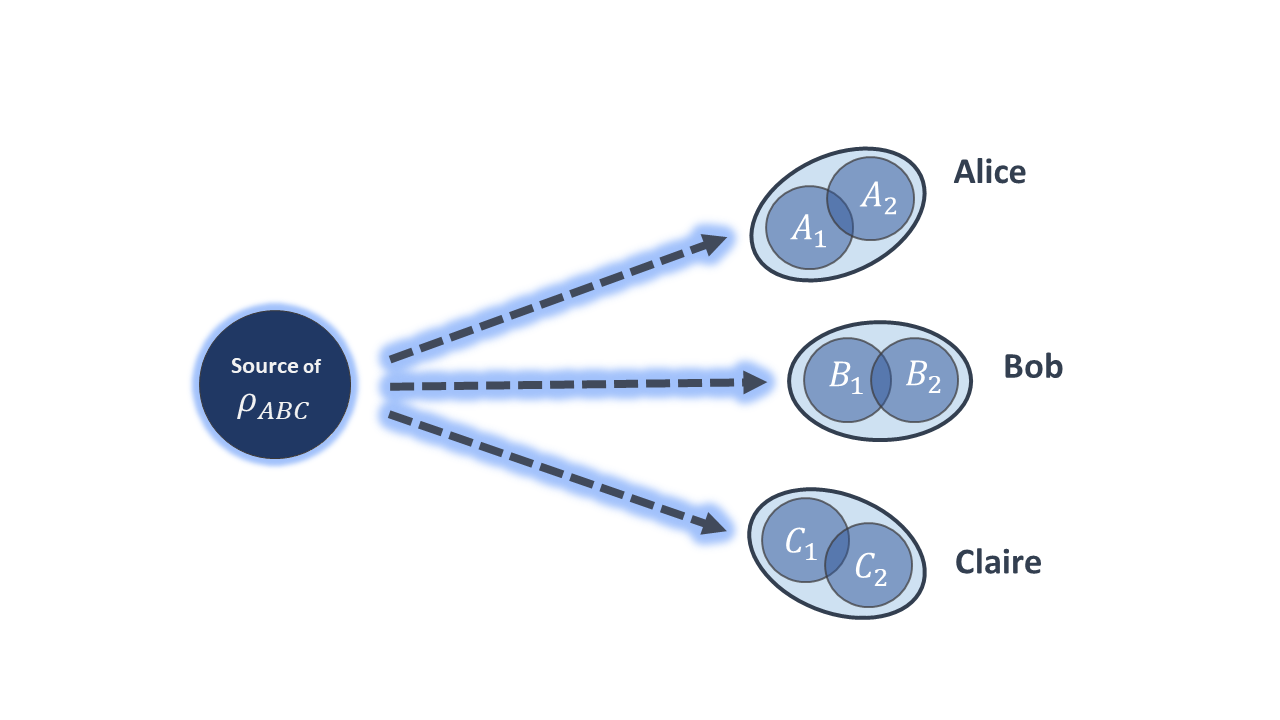}
        \vspace*{-1.5mm}\caption{\textbf{Multi-copy} \textbf{GME activation.
        } 
       \textbf{(a)} Separability structure for tripartite systems: Fully separable states $\rho_{A|B|C}$ (dark blue) form a (convex) subset $\mathcal{S}_3$ of the intersection of the three (convex) sets of partition-separable states, $\rho_{AB|C},$ $\rho_{AC|B}$, and $\rho_{BC|A}$ (three light-blue regions). The convex hull of all partition-separable states forms the set $\mathcal{S}_2$ of biseparable states (all blue and red regions). The FIB states lie in the red region. All other fully inseparable states lie in the grey area outside of $\mathcal{S}_2$ and are GME. \textbf{(b)} Multi-copy GME activation{\textemdash}two-copy case in a tripartite scenario: two copies of a biseparable state may be GME with respect to the partition $A_1A_2|B_1B_2|C_1C_2$.}
    \label{fig:gme_graph}
    \end{figure}

The sets $\mathcal{S}_k$ formed by all states that are (at least) $k$-separable form a hierarchy of nested convex sets, $\mathcal{S}_N\subseteq \ldots\subseteq \mathcal{S}_k\subseteq \ldots \mathcal{S}_3 \subseteq \mathcal{S}_2$. Here it is crucial to note that the set $\mathcal{S}_2$ of biseparable states is the convex hull of all partition-separable states. 
As such, $\mathcal{S}_2$ contains some states that are fully inseparable and thus multipartite entangled. Yet, only states that are not (at least) biseparable, and which are hence outside of the set $\mathcal{S}_2$, are genuinely $N$-partite entangled or \emph{genuinely multipartite entangled}.  
For reviews, see, e.g., Refs.~\cite{GuehneToth2009,FriisVitaglianoMalikHuber2019} or~\cite[Chapter~18]{BertlmannFriis2023}. 

In this paper, we will pay special attention to the states that belong to the set $\mathcal{S}_2$ but which do not belong to any set of partition-separable states; we will call these \textit{fully inseparable biseparable} (FIB) states. These are the states that are potentially GME activatable, and, indeed, it was shown in Ref.~\cite{PalazuelosDeVicente2022} that all such states in finite-dimensional Hilbert spaces are GME activatable. That is, for any FIB state $\rho\subtiny{0}{0}{ABC\ldots}$ in a finite-dimensional Hilbert space there exists a $k\geq2$ such that $\rho\subtiny{0}{0}{ABC\ldots}^{\otimes k}=\rho\subtiny{0}{0}{A_{1}B_{1}C_{1}\ldots}\otimes\cdots\otimes\rho\subtiny{0}{0}{A_{k}B_{k}C_{k}\ldots}$ is GME with respect to the partition $A_{1}\ldots A_{k}|B_{1}\ldots B_{k}|C_{1}\ldots C_{k}|\ldots$\,. A graphical representation of the state space of three parties, along with an illustration of the
partition for two-copy ($k=2$) three-partite GME activation, is shown in Fig.~\ref{fig:gme_graph}.

\subsection{Continuous-variable systems}

In the following, we consider infinite-dimensional quantum systems comprising $N$ modes (e.g., of the electromagnetic field). To each mode labelled $j=1,\ldots,N$, one associates operators ${a}_j=\tfrac{1}{\sqrt{2}}\bigl({x}_j+i{p}_j\bigr)$ and ${a}^\dagger_j=\tfrac{1}{\sqrt{2}}\bigl({x}_j-i{p}_j\bigr)$, where ${x}_j$ and ${p}_j$ are quadrature operators satisfying $ [{x}_j,{p}_k] = i\delta_{jk},\ \ \ \left[{x}_j,{x}_k\right] =\left[{p}_j,{p}_k\right] =0\,$. The latter can be arranged into a vector $\mathbf{{\mathbf{r}}} = \left({x}_1,{p}_1,\ldots,{x}_N,{p}_N\right)^\mathrm{T}$. The properties of CV systems can be described by the Wigner function,
\begin{align}\label{Wigner}
W(\mathbf{x},\mathbf{p})[{\rho}] = \tfrac{1}{\left(2\pi\right)^N}\!\!\int \!\!\mathrm{d}^{N}\!\mathbf{x'}\,e^{i\mathbf{x'}\cdot\mathbf{p}}\bra{\mathbf{x}\hspace*{-1pt}-\hspace*{-1pt}\tfrac{\mathbf{x'}}{2}}{\rho}\ket{\mathbf{x}\hspace*{-1pt}+\hspace*{-1pt}\tfrac{\mathbf{x'}}{2}},
\end{align}
with $\ket{\mathbf{x}\pm\tfrac{\mathbf{x'}}{2}} = \bigotimes^N_{i=1}\ket{x_i\pm\tfrac{x'_i}{2}}$, where $\ket{x_i}$ are eigenstates of $x_i$. An important family of CV states are so-called \emph{Gaussian states}, whose Wigner function is Gaussian and reduces to
\begin{align}\label{WigGauss}
    W(\mathbf{r})=\frac{e^{-(\mathbf{r}-\mathbf{d})^\mathrm{T}\gamma^{-1}(\mathbf{r}-\mathbf{d})}}{\pi^N\sqrt{\mathrm{det}(\gamma)}}.
\end{align}
Any Gaussian state thus is fully determined by its vector of first moments $\mathbf{d}$ with elements $d_i=\left\langle {r}_i\right\rangle=\mathrm{Tr}({\rho}\,{r}_i)$ and its CM $\gamma$ with components $\gamma_{ij}=\left\langle {r}_i {r}_j+ {r}_j {r}_i\right\rangle-2\left\langle {r}_i\right\rangle\left\langle {r}_j\right\rangle$. 
Since the first moments $d_i$ can be set to zero via local displacements, correlations in Gaussian states are fully described by their CM~$\gamma$. 
States whose Wigner function is not of the form of Eq.~(\ref{WigGauss}) are called non-Gaussian states. This is the weakest condition for non-Gaussianity and serves as the definition we adopt in this work. In contrast, some tasks require stronger conditions, for example, so-called quantum non-Gaussianity requires a Wigner function that cannot even be expressed as a convex combination of Gaussian Wigner functions~\cite{FilipMista2011}. In some cases, a negative Wigner function is required~\cite{KenfackZyczkowski2004}. Both cases thus refer to subsets of states that we call non-Gaussian here. For a review of non-Gaussian states we refer to Ref.~\cite{Walschaers2021}. For reviews of CV systems for quantum-information processing, see, e.g., Refs.~\cite{WeedbrookPirandolaGarciaPatronCerfRalphShapiroLloyd2012, AdessoRagyLee2014}.

\section{GME activation in infinite dimensions}\label{sc:GMEactivationinfinitedimension}

In this section, we extend the result of GME activatability of all FIB states in finite dimensions from Ref.~\cite{PalazuelosDeVicente2022} to all FIB states in infinite-dimensional systems. To this end, we leverage results from Ref.~\cite{SperlingVogel2009}, which guarantee the existence of local projections that map states with bipartite entanglement in continuous variables (CVs) to states with bipartite entanglement in finite dimensions. Let us begin by phrasing the pertinent results (Theorem 2 and Lemma 1) of Ref.~\cite{SperlingVogel2009} in the following way suitable to our discussion here:\\[-2mm]

For every $N$-partite CV state $\rho$ with the $N$ parties divided into two groups $A$ and $B$ such that $\rho$ is entangled with respect to a bipartition $A|B$ of the total Hilbert space $\mathcal{H}=\mathcal{H}_{A}\otimes\mathcal{H}_{B}$ there exists a local projection $P_{d}=P_{d}\suptiny{1}{0}{A}\otimes P_{d}\suptiny{1}{0}{B}$, where $P_{d}^{j}=\sum_{k=1}^{d}|e_{j,k}\rangle\langle e_{j,k}|$, $j=A,B$, where $\{|e_{j,k}\rangle\}_{k\in\mathbb{N}_{0}}$ is an orthonormal basis in the state space 
$\mathcal{H}_{j}$ of subsystem $j$, projecting into a subspace $\mathcal{H}_{d}$ of $\mathcal{H}$ with finite dimension $d$, with local projectors $P_{d}\suptiny{1}{0}{A}$ and $P_{d}\suptiny{1}{0}{B}$ into (finite-dimensional) subspaces of $\mathcal{H}_{A}$ and $\mathcal{H}_{B}$, respectively, such that:
\vspace*{-0.5mm}
\begin{enumerate}
        \item{\label{first} The state $\rho_{d}=P_{d}\,\rho\,P_{d}/\tr(P_{d}\,\rho\,P_{d})$ is entangled with respect to the same bipartition as~$\rho$.}\\[-5mm]
        \item{\label{second} 
        For all $d^{\,\prime}\geq d$ one may define local projectors $P_{d^{\,\prime}}$ such that $\rho_{d^{\,\prime}}=P_{d^{\,\prime}}\,\rho\,P_{d^{\,\prime}}/\tr(P_{d^{\,\prime}}\,\rho\,P_{d^{\,\prime}})$ is entangled with respect to the same bipartition as $\rho$, and $\rho_{d'}\rightarrow \rho$ (in trace norm) as $d'\rightarrow\infty$~\cite[Lemma 1]{SperlingVogel2009}.}
        \end{enumerate}
Based on these statements, we now formulate the following corollary:
\\[-2mm]

\noindent\textbf{Corollary 1} \textit{For every CV FIB state $\rho$ there exists~a local projection $P_{d}$ into a finite-dimensional Hilbert $\mathcal{H}_{d}$ with dimension $d=\dim(\mathcal{H}_{d})<\infty$, such that $\rho$ is projected onto~a finite-dimensional 
FIB state 
\begin{align}\label{finBNPS}
    \rho_{d} &=\,\frac{P_{d}\,\rho \,P_{d}}{\tr(P_{d}\,\rho \,P_{d})}\,.
\end{align}}

The two points above,~\ref{first} and~\ref{second}, guarantee that the local projections can be chosen so that entanglement with respect to any fixed bipartition is preserved. However, a priori it is not clear that one may choose a projection that simultaneously preserves entanglement with respect to all pertinent bipartitions, which is necessary to ensure that all CV FIB states are projected onto FIB states in finite dimensions.

To claim this subtle point, we argue that properties of the local projection $P_d$ directly imply that among the projections preserving bipartite entanglement with respect to particular bipartitions of a given state, the one projecting into the subspace with the largest dimension preserves bipartite entanglement in all other bipartitions while it also cannot create GME, which proves Corollary 1 (see Appendix~\ref{SM:4} for more details).

Finally, all that is left to note is that, according to Ref.~\cite{PalazuelosDeVicente2022}, all FIB states in finite dimensions are GME activatable. And since 
\begin{align}\label{eq:kfoldproduct}
    \rho_d^{\otimes k}=P_d^{\otimes k}\rho^{\otimes k}P_d^{\otimes k},
\end{align}
then GME activitability of $\rho_d$ together with locality of $P_d$ imply that also the original FIB state $\rho$ in infinite dimensions must be GME activatable.

\section{GME activation for non-Gaussian states}

Now we turn to the demonstration of GME activation for non-Gaussian states. Since non-Gaussian states that have an overlap with only finitely many Fock states can be represented completely on a finite-dimensional Hilbert space, their GME activatability follows trivially from the results of Ref.~\cite{PalazuelosDeVicente2022}. Thus, we will present an example of GME activation for states that have non-zero overlap with infinitely many Fock states. To this end, we construct a one-parameter family of three-mode non-Gaussian states with this property by considering convex combinations of the tensor product of two-mode squeezed vacuum (TMSV) states $\rho\suptiny{0}{0}{\rm TMSV}=(1-\lambda^2)\sum_{m,m'=0}\suptiny{0}{0}\infty\lambda^{m+m'}|mm\rangle\langle m'm'|$ with $\lambda=\tanh{r}$, and $n$-excitation Fock states $\ket{n}$ in the third mode. In this way, we obtain fully symmetric (FS) states
\begin{small}
\begin{align}\label{FSmain}
\begin{split}
    \rho\suptiny{0}{0}{\rm FS}\subtiny{0}{0}{ABC}=&\tfrac{1}{3}\bigl(\rho\suptiny{0}{0}{\rm TMSV}\subtiny{0}{0}{AB}\!\otimes\ket{n}\!\!\bra{n}\subtiny{0}{0}{C}\! 
    \,+\, \rho\suptiny{0}{0}{\rm TMSV}\subtiny{0}{0}{AC}\!\otimes\ket{n}\!\!\bra{n}\subtiny{0}{0}{B}\!\\[1.5mm]
    &+\,\ket{n}\!\!\bra{n}\subtiny{0}{0}{A}\!\otimes\rho\suptiny{0}{0}{\rm TMSV}\subtiny{0}{0}{BC}\bigr). 
\end{split}
\end{align}
\end{small}

{\noindent}These states are biseparable by construction but entangled with respect to all three bipartitions, and hence fully inseparable for all non-zero values of the squeezing parameter $r\neq 0$ and all excitation numbers~$n$. This can be seen by noting that the two-qubit states are obtained by tracing out any single mode, e.g., $C$, and locally projecting the remaining two modes into the subspace spanned by any two local Fock-state pairs $\{\ket{k}\subtiny{0}{0}{A}, \ket{k'}\subtiny{0}{0}{A}\}$ and $\{\ket{k}\subtiny{0}{0}{B}, \ket{k'}\subtiny{0}{0}{B}\}$ for $k,k'\neq n$ and $k\neq k'$ are entangled, cf.~Appendix~\ref{SM:1.1}.

For investigating multipartite entanglement in CV systems several methods are available (see, e.g.,~\cite{Simon2000, DuanGiedkeCiracZoller2000, ShchukinVogel2006, SperlingVogel2013, TehReid2014, ShchukinLoock2014, ShchukinLoock2015, ZhangBarralZhangXiaoBencheikh2023}). Here, we use a special case of a $k$-separability criterion~\cite{GabrielHiesmayrHuber2010}: 
Every $k$-separable $N$-partite state $\rho$ satisfies
\begin{small}
\begin{align}\label{cr:Gabrielmain}
    \sqrt{\bra{\phi}\rho^{\otimes 2}P_{\mathrm{tot}}\ket{\phi}}
    \leq
    \sum_{\{M\}}\Bigl(\prod_{i=1}^k\bra{\phi}P^\dagger_{M(i)}\rho^{\otimes 2}P_{M(i)}\ket{\phi}\Bigr)\suptiny{0}{0}{\tfrac{1}{2k}},
\end{align}
\end{small}

{\noindent}for every fully separable $2N$-partite state $\ket{\phi}=\bigotimes_{i=1}^{2N}\ket{\phi\subtiny{0}{0}{i}}$, where $P_{M(i)}$ are permutation operators exchanging the two copies of all subsystems contained in the $i$-th subset of the partition $M$, $P_{\mathrm{tot}}$ exchanges the two copies entirely, and the sum runs over all possible partitions $M$ of the considered system into $k$ subsystems. Violating the Ineq.~(\ref{cr:Gabrielmain}) for $k=2$ thus detects genuine $N$-partite entanglement. 

We now employ this criterion for $k=2$, i.e., to check if the state given by two copies of $\rho\suptiny{0}{0}{\rm FS}\subtiny{0}{0}{ABC}$ is GME. Thus, the state $\rho$ in Ineq.~(\ref{cr:Gabrielmain}) is $\rho=\rho\suptiny{0}{0}{\rm FS}\subtiny{0}{0}{A_{1}B_{1}C_{1}}\otimes\rho\suptiny{0}{0}{\rm FS}\subtiny{0}{0}{A_{2}B_{2}C_{2}}$ and we pick $\ket{\phi}$ to be the fully separable state
\begin{align}
\begin{split}
    \ket{\phi}  &=\ket{n00}\subtiny{0}{0}{A_{1}B_{1}C_{1}}\otimes\ket{0n0}\subtiny{0}{0}{A_{2}B_{2}C_{2}}\\[1mm]
    &\ \ \ \ \ \otimes\ket{n11}\subtiny{0}{0}{A' _{1}B'_{1}C'_{1}}
    \otimes\ket{1n1}\subtiny{0}{0}{A' _{2}B'_{2}C'_{2}}\,.
\end{split}
\end{align}
For this choice, the left-hand side of Ineq.~(\ref{cr:Gabrielmain}) evaluates to
\begin{align}
\begin{split}
    &|\bra{n00}\rho\suptiny{0}{0}{\rm FS}\subtiny{0}{0}{ABC}\ket{n11}|\times
    |\bra{0n0}\rho\suptiny{0}{0}{\rm FS}\subtiny{0}{0}{ABC}\ket{1n1}|\\[1mm]
    &=\,
    \tfrac{1}{9}(1-\lambda^{2})^{2}\lambda^{2}\,,
\end{split}
\end{align}
whereas each term on the right-hand side is proportional to $|\bra{0n1}\rho\suptiny{0}{0}{\rm FS}\subtiny{0}{0}{ABC}\ket{0n1}|=0$ or $|\bra{n01}\rho\suptiny{0}{0}{\rm FS}\subtiny{0}{0}{ABC}\ket{n01}|=0$ (see Appendix~\ref{SM:1.2} for more details). The inequality is violated for all non-zero values of $r$. The two-copy state is GME, even though a single copy is biseparable, which explicitly illustrates that GME activation is possible in infinite-dimensional systems for non-Gaussian states. 

\section{GME activation for Gaussian states}

We now turn to the characterization of the multipartite entanglement structure for Gaussian states. Since the correlations of the latter are fully captured by their second moments, the CM offers itself for this task. Indeed, it has been shown~\cite{WernerWolf2001, AndersPhDThesis2003} that a Gaussian state $\rho$ with CM $\gamma$ is fully separable with respect to a partition into $N$ subsystems (of one or more modes each) if and only if there exist CMs $\gamma\suptiny{0}{0}{(i)}$ for $i=1,\ldots,N$ corresponding to these $N$ subsystems such that $\gamma-\gamma\suptiny{0}{0}{(1)}\oplus\cdots\oplus\gamma\suptiny{0}{0}{(N)}\geq0$. For arbitrary (not necessarily Gaussian) states that are fully separable, such a decomposition also exists, but the existence of such a decomposition generally does not imply full separability. A generalization that we call the \emph{CM biseparability criterion} was given in Ref.~\cite{HyllusEisert2006}: 
For any biseparable state with CM $\gamma\subtiny{0}{0}{\mathrm{BS}}$ there exist block-diagonal CMs $\gamma\subtiny{0}{0}{M(i)}$ corresponding to the partition $M(i)$ along with a probability distribution $\{p_i\}$ such that 
\begin{align}\label{bisepCMmain}
\gamma\subtiny{0}{0}{\mathrm{BS}}-\sum_i \,p_i\,\gamma\subtiny{0}{0}{M(i)}\,\geq\,0\,.
\end{align}
If no such convex decomposition into CMs $\gamma\subtiny{0}{0}{M(i)}$ exists, one can hence conclude that the state under consideration must be GME. 

However, as we shall show now, this criterion cannot be used to detect GME activation for identical copies: 
If a CM $\gamma\subtiny{0}{0}{\mathrm{BS}}$ corresponding to a state $\rho$ satisfies the condition~(\ref{bisepCMmain}), then so does the CM $\bigoplus_{n=1}^{k}\gamma\subtiny{0}{0}{\mathrm{BS}}$ corresponding to the $k$-copy state $\rho^{\otimes k}$. 
To prove this, we note that if for a given CM $\gamma\subtiny{0}{0}{\mathrm{BS}}$ the ensemble $\{(p_i,\gamma\subtiny{0}{0}{M(i)})\}_{i}$ is such that $\Delta\gamma:=\gamma\subtiny{0}{0}{\mathrm{BS}}-\sum_i \,p_i\,\gamma\subtiny{0}{0}{M(i)}\geq0$, then the ensemble $\{(p_i,\bigoplus_{n=1}^{k}\gamma\subtiny{0}{0}{M(i)}\suptiny{0}{0}{(n)})\}_{i}$ with $\gamma\subtiny{0}{0}{M(i)}\suptiny{0}{0}{(n)}=\gamma\subtiny{0}{0}{M(i)}\,\forall\,n$ satisfies
\begin{align}\label{eq:resultmain}
\begin{split}
    &\bigoplus_{n=1}^{k}\gamma\subtiny{0}{0}{\mathrm{BS}}
    -
    \sum_i \,p_i\,\bigoplus_{n=1}^{k}\gamma\subtiny{0}{0}{M(i)}\suptiny{0}{0}{(n)}
    = \\
    &=\bigoplus_{n=1}^{k}\bigl(\gamma\subtiny{0}{0}{\mathrm{BS}}-
    \sum_i \,p_i\,\gamma\subtiny{0}{0}{M(i)}\bigr)\,\geq\,0
    \,,
\end{split}
\end{align}
since the left-hand side is block-diagonal and each block is identical to a positive semi-definite matrix $\Delta\gamma\geq0$. 
While this result means that the CM biseparability criterion cannot be used to detect potential GME activation for identical copies of a given state, it may still succeed in detecting GME for pairs of two (or more) different FIB Gaussian states with CMs $\gamma$ and $\tilde{\gamma}$, respectively, as long as the states do not admit `biseparable' CM decompositions $\{(p_i,\gamma\subtiny{0}{0}{M(i)})\}_{i}$ and $\{(q_i,\tilde{\gamma}\subtiny{0}{0}{M(i)})\}_{i}$ with $p_i=q_i\ \forall\, i\,$. 
In Appendix~\ref{SM:2}, we present examples of such a GME activation from pairs of different Gaussian states. 

The perhaps more pressing question concerning the result in Ineq.~(\ref{eq:resultmain}) is whether it permits GME activation for identical copies of Gaussian states at all. 
That is, if the CM biseparability criterion Ineq.~(\ref{bisepCMmain}) was necessary and sufficient for biseparability of Gaussian states in analogy to the criterion for (full) separability~\cite{WernerWolf2001, AndersPhDThesis2003}, then no Gaussian GME activation would be possible. However, we will show next that satisfying the CM biseparability criterion (\ref{bisepCMmain}) is not sufficient for the biseparability of Gaussian states. For this purpose, we focus on an example of a three-mode Gaussian state with CM 
\begin{align}\label{testCMmain}
    \gamma\subtiny{0}{0}{ABC}
    =\tfrac{1}{3}\bigl(
    \gamma\subtiny{0}{0}{AB}\suptiny{0}{0}{\mathrm{TMSV}} \hspace*{-1pt}\oplus\hspace*{-1pt} \mathds{1}\subtiny{0}{0}{C}
    \hspace*{-1pt}+\hspace*{-1pt}\gamma\subtiny{0}{0}{BC}\suptiny{0}{0}{\mathrm{TMSV}} \hspace*{-1pt}\oplus\hspace*{-1pt} \mathds{1}\subtiny{0}{0}{A}
    \hspace*{-1pt}+\hspace*{-1pt}\gamma\subtiny{0}{0}{AC}\suptiny{0}{0}{\mathrm{TMSV}} \hspace*{-1pt}\oplus\hspace*{-1pt} \mathds{1}\subtiny{0}{0}{B}\bigr),
\end{align}
where
\begin{align}
    \gamma\suptiny{0}{0}{\mathrm{TMSV}} = 
    \left(\begin{array}{cc}
       \cosh(2r)\mathds{1}  & \sinh(2r)Z \\
        \sinh(2r)Z & \cosh(2r)\mathds{1}
    \end{array}
    \right)
\end{align}
is the CM of a TMSV state, $Z=\operatorname{diag}\{1,-1\}$ is the usual third Pauli matrix, and $\mathds{1}$ is the CM of the single-mode vacuum state. One observes that this is the same CM as that of the non-Gaussian state $\rho\suptiny{0}{0}{\rm FS}\subtiny{0}{0}{ABC}$ in Eq.~(\ref{FSmain}) for $\ket{n}=\ket{0}$, but here we use it to define a Gaussian state $\rho\suptiny{0}{0}{\rm G}\subtiny{0}{0}{ABC}$ with zero first moments. Moreover, we note that $\gamma\subtiny{0}{0}{ABC}$ satisfies the CM biseparability criterion by construction.  

Nevertheless, we find that the state is certainly GME for the parameter range $0<r<r_{0}$ with $r_{0}\approx0.575584$. Between $r_{0}$ and $r_{1}=\tfrac{1}{2}\operatorname{arcosh}([7 + 2 \sqrt{31}]/3)\approx1.24275$ the three-mode state is fully inseparable and GME activatable (or potentially already GME at the single-copy level). For $r>r_{1}$, the state is partition separable and thus certainly not GME activatable. Let us now discuss how to obtain these values. For full inseparability, the threshold value $r_{1}$ is obtained directly from the CM, where the PPT criterion provides a necessary and sufficient criterion for separability of $1$ vs. $N$-mode Gaussian states~\cite{WernerWolf2001}, as we discuss in more detail in 
Appendix~\ref{SM:3.1}. 

For the detection of GME we employ different methods. Up to the value $r_{0}^{\prime}=0.284839$, we detect GME by employing another witness inequality satisfied by all biseparable states $\rho\suptiny{0}{0}{\rm BS}\subtiny{0}{0}{ABC}$. This witness stated fully and proven in 
Appendix~\ref{SM:3.2}, is a generalization of a witness that first appeared in Ref.~\cite{FriisHuberFuentesBruschi2012}, constructed using techniques similar to those in Refs.~\cite{GabrielHiesmayrHuber2010,HuberMintertGabrielHiesmayr2010}. Taking into account the symmetry of the state $\rho\suptiny{0}{0}{\rm G}\subtiny{0}{0}{ABC}$, the inequality reduces to
\begin{align}
    &\sqrt{3}\,
    |\bra{000}\rho\suptiny{0}{0}{\rm G}\subtiny{0}{0}{ABC}\ket{011}|
    \label{ineq FS state}\\[1mm]
    &\ \leq\sqrt{\bra{000}\rho\suptiny{0}{0}{\rm G}\subtiny{0}{0}{ABC}\ket{000}\,
    \bra{011}\rho\suptiny{0}{0}{\rm G}\subtiny{0}{0}{ABC}\ket{011}
    }\nonumber\\[1mm]
    &\ \ \ \ \ +\,\sqrt{3}\,\bra{001}\rho\suptiny{0}{0}{\rm G}\subtiny{0}{0}{ABC}\ket{001}\,.
    \nonumber
\end{align}
We calculate the relevant density-matrix elements of the Gaussian state $\rho\suptiny{0}{0}{\rm G}\subtiny{0}{0}{ABC}$ from its CM in Eq.~(\ref{testCMmain}).
As explained in detail in 
Appendix~\ref{SM:3.3}, this leads to a violation of the Ineq.~(\ref{ineq FS state}) in the parameter range $0<r<r_{0}^{\prime}$. 
Thus, we conclude that the CM biseparability criterion cannot be sufficient for biseparability even for Gaussian states.  

What we can further conclude from the calculated density-matrix elements is that the state is GME (at least) up to the larger value $r_{0}\approx0.575584$, and GME activatable for all values of $r$ between $r_0$ and $r_1$. But we do not know if it is GME on the single-copy level between $r_0$ and $r_1$. We conclude this via a local filtering operation $\Lambda$ that maps the three-mode state $\rho\suptiny{0}{0}{\rm G}\subtiny{0}{0}{ABC}$ to a three-qubit state $\rho\suptiny{0}{0}{\rm QB}\subtiny{0}{0}{ABC}=\Lambda[\rho\suptiny{0}{0}{\rm G}\subtiny{0}{0}{ABC}]$ by projecting the former into the subspace spanned by the Fock states with at most one excitation in each mode. 
This operation cannot create entanglement. For $0<r<r_{0}$, the three-qubit state is detected as GME by a fully decomposable witness~\cite{JungnitschMoroderGuehne2011a}. 
For $0<r<r_{1}$ we find that $\rho\suptiny{0}{0}{\rm QB}\subtiny{0}{0}{ABC}$ is detected as bipartite entangled by the PPT criterion~\cite{Peres1996,HorodeckiMPR1996,Horodecki1997}. From the symmetry of the state, we can thus infer that $\rho\suptiny{0}{0}{\rm QB}\subtiny{0}{0}{ABC}$, and hence $\rho\suptiny{0}{0}{\rm G}\subtiny{0}{0}{ABC}$ must be fully inseparable for $0<r<r_{1}$. Moreover, from Refs.~\cite{PalazuelosDeVicente2022,SperlingVogel2009}, it follows that there is some $k\geq2$ such that $(\rho\suptiny{0}{0}{\rm QB}\subtiny{0}{0}{ABC})^{\otimes k}$ is GME (if ${\rho\suptiny{0}{0}{\rm QB}\subtiny{0}{0}{ABC}}$ is not already GME) in the same region, and since $(\rho\suptiny{0}{0}{\rm QB}\subtiny{0}{0}{ABC})^{\otimes k}=\Lambda^{\otimes k}[(\rho\suptiny{0}{0}{\rm G}\subtiny{0}{0}{ABC})^{\otimes k}]$, also ${\rho\suptiny{0}{0}{\rm G}\subtiny{0}{0}{ABC}}^{\otimes k}$ must be at least GME activatable. 

A corollary of our results is that a Gaussian state may have the same first and second moments as a biseparable non-Gaussian state [see Eq.~(\ref{testCMmain}) in Appendix~\ref{SM:2}], yet itself be GME. 
Thus, no GME criterion valid for all states that is based solely on first and second moments of a state can detect such Gaussian-state GME. Any detection of GME must hence rely on higher statistical moments, even if those are themselves functions only of the first and second moments if the state is Gaussian. 

\section{Conclusion and outlook}

We showed that the activation of GME from multiple identical copies of the state is also possible in infinite dimensions. We demonstrated this specifically for a family of non-Gaussian states with non-zero overlap with infinitely many Fock states. We then investigated the GME activatability of Gaussian states. However, as we showed, this matter is complicated by the fact that the CM biseparability criterion is not sufficient for biseparability, even for Gaussian states. In particular, we demonstrated that Gaussian states satisfying the CM biseparability criterion could be GME. Interestingly, this is the case even though satisfying the CM biseparability criterion implies that the corresponding Gaussian states have the same first and second moments as biseparable non-Gaussian states.

At the same time, our results leave us without an easily verifiable sufficient criterion for the biseparability of Gaussian states if no explicit decomposition into a convex sum of partition-separable states is given. We thus lack a tool to conclusively determine if GME-activatable Gaussian states are not already GME on the single-copy level to begin with. In other words, we are not aware of any example of a fully inseparable yet provably biseparable (red area in Fig.~\ref{fig:gme_graph}) Gaussian state. We leave the development of suitable techniques to address this question for future research.

\section*{Acknowledgments}
We thank Nicky Kai Hong Li for useful comments on a previous version of this manuscript. 
K.B. and N.F. acknowledge support from the Austrian Science Fund (FWF) through the project P 36478-N funded by the European Union - NextGenerationEU. 
O.L. and L.M. acknowledge project 8C22002 (CVStar) of MEYS of the Czech Republic, which has received funding from the European Union's Horizon 2020 Research and Innovation Programme under Grant Agreement no. 731473 and 101017733.
E.A. acknowledges funding from the Austrian Science Fund (FWF) through the Lise Meitner-Programm project M3151 and the Elise Richter project V1037. 
N.F. also acknowledges funding from the Austrian Federal Ministry of Education, Science, and Research via the Austrian Research Promotion Agency (FFG)  
through the flagship project FO999897481 (HPQC), the project FO999914030 (MUSIQ), and the project FO999921407 (HDcode) funded by the European Union{\textemdash}NextGenerationEU. 
This research was funded in whole or in part by the Austrian Science Fund (FWF) [\href{https://doi.org/10.55776/P36478}{10.55776/P36478}]. For open access purposes, the author has applied a CC BY public copyright license to any author accepted manuscript version arising from this submission.


%

\onecolumn
\section*{Appendix}
In the appendix, we present additional details and explicit calculations supporting our results.
The appendix is structured as follows: In Sec.~\ref{SM:1} we present additional details on the GME activation for non-Gaussian states.
In Sec.~\ref{SM:2} we provide a detailed description of GME activation for non-identical Gaussian states.
Section~\ref{SM:3} shows that Gaussian states satisfying the CM biseparability criterion can be GME. Finally, Sec.~\ref{SM:4} provides a proof that all FIB states are GME activatable. 
To provide a better overview, Table~\ref{tab:term} summarizes the terminology related to multipartite entanglement that we use throughout.

\appendix

\begin{table*}[b!]
\centering
\caption{\label{tab:term}Terminology: Separability structure in multipartite scenarios~\cite{YamasakiMorelliMiethlingerBavarescoFriisHuber2022}.}

\begin{tabular}{@{}ll@{}}
\toprule
Term & Meaning\\
\midrule
\vspace*{3mm}
$k$-separable &\parbox{12cm}{convex combination of pure states, each of which is a product of at least $k$ projectors}\\\vspace*{3mm}

fully separable & synonymous with $N$-separable for $N$ parties\\\vspace*{3mm}

biseparable & synonymous with $2$-separable\\\vspace*{3mm}

partition-separable & \parbox{12cm}{\raggedright separable for a specific fixed partition of the multipartite Hilbert space, i.e., a convex combination of projectors, each of which is a product with respect to the same partition into subsystems}\\\vspace*{3mm}

fully inseparable & entangled across all bipartitions\\\vspace*{3mm}

genuine multipartite entangled & non-biseparable\\

\end{tabular}
\end{table*}

\section{Additional details on the GME activation for non-Gaussian states}\label{SM:1}

\subsection{Full inseparability of biseparable non-Gaussian states}\label{SM:1.1}

We begin by showing in more detail that the members of the one-parameter family of fully symmetric (FS) non-Gaussian states
\begin{align}\label{FS}
\scalebox{0.95}{\mbox{\ensuremath{
    \rho\suptiny{0}{0}{\rm FS}\subtiny{0}{0}{ABC}
    = \tfrac{1}{3}\bigl(\rho\suptiny{0}{0}{\rm TMSV}\subtiny{0}{0}{AB}\!\otimes\ket{n}\!\!\bra{n}\subtiny{0}{0}{C}\! 
    + \rho\suptiny{0}{0}{\rm TMSV}\subtiny{0}{0}{AC}\!\otimes\ket{n}\!\!\bra{n}\subtiny{0}{0}{B}\!
    +\ket{n}\!\!\bra{n}\subtiny{0}{0}{A}\!\otimes\rho\suptiny{0}{0}{\rm TMSV}\subtiny{0}{0}{BC}\bigr),
}}}    
\end{align}
where 
\begin{align}
    \rho\suptiny{0}{0}{\rm TMSV} 
    &= \,(1-\lambda^2) \sum\limits_{m,m'=0}^\infty\lambda^{m+m'}\ket{mm}\!\!\bra{m'm'}
\end{align}
is two-mode squeezed vacuum (TMSV) state with $\lambda=\tanh{r}$, and $\ket{n}$ is $n$-excitation Fock state, are FIB states. Biseparability is ensured by construction since the states are (equally weighted) mixtures of product states.\\[-3mm]

For $r=0$, the state is a convex mixture of products of the vacuum and Fock states and is thus separable. 
For all non-zero values of $r$ we will now show that the states $\rho\suptiny{0}{0}{\rm FS}\subtiny{0}{0}{ABC}$ are entangled across all bipartitions. 
To do this, we note that the symmetry of the state with respect to the exchange of the modes means that it is sufficient to show that the state is entangled for any fixed bipartition, e.g., $A|BC$. 
We then trace out the third mode, $C$, an operation that cannot create entanglement between $A$ and $B$ where none was present before, and we are left with the reduced state
\begin{align}
    \rho\suptiny{0}{0}{\rm FS}\subtiny{0}{0}{AB} \nonumber   
    &= \tr\subtiny{0}{0}{C}(\rho\suptiny{0}{0}{\rm FS}\subtiny{0}{0}{ABC})
    = 
    \tfrac{1}{3}\bigl(
    \rho\suptiny{0}{0}{\rm TMSV}\subtiny{0}{0}{AB}\!\!
    +
    \rho\suptiny{0}{0}{\rm Th}\subtiny{0}{0}{A}\otimes\ket{n}\!\!\bra{n}\subtiny{0}{0}{B}
    \!+\ket{n}\!\!\bra{n}\subtiny{0}{0}{A}\!\otimes\rho\suptiny{0}{0}{\rm Th}\subtiny{0}{0}{B}\bigr),
\end{align}
where $\rho\suptiny{0}{0}{\rm Th}=(1-\lambda^2) \sum_{m=0}^\infty \lambda^{2m} \ket{m}\!\!\bra{m}$ is a single-mode thermal state. 
Now we can choose any pair of excitation numbers different from $n$, let us label them $k$ and $k'$, and project into the subspace spanned by the product states $\ket{i,j}\subtiny{0}{0}{AB}$ for $i,j=k,k'\neq n$. 
This is a local map that also cannot create entanglement. 
After normalization, one obtains the two-qubit density operator
\begin{align}
    \rho\suptiny{0}{0}{\rm QB}\subtiny{0}{0}{AB}
    &=\,
    \tfrac{1}{\lambda^{2k}+\lambda^{2k'}}\sum\limits_{m,m'=k,k'}\lambda^{m+m'}\ket{mm}\!\!\bra{m'm'}\,.
\end{align}
This is a pure two-qubit state $\rho\suptiny{0}{0}{\rm QB}\subtiny{0}{0}{AB}=\ket{\psi_{kk' }}\!\!\bra{\psi_{kk' }}$ with $\ket{\psi_{kk' }}=(\lambda^k \ket{kk}+\lambda^{k'}\ket{k'k'})/\sqrt{\lambda^{2k}+\lambda^{2k'}}$ that is not a product state, and hence entangled, for all $r\neq0$.\\[-2mm]

Although investigating inseparability of non-Gaussian mixed states in an infinite-dimensional Hilbert space through the negativity of their partial transpose is in general challenging, in the case of $\rho\suptiny{0}{0}{\rm FS}\subtiny{0}{0}{ABC}$ from Eq.~(\ref{FS}) with $n=0$, one can easily obtain the eigenvalues of the partial tranpose analytically. For this purpose, let us write the partially transposed state $\left(\rho\suptiny{0}{0}{\rm FS}\subtiny{0}{0}{ABC}\right)\suptiny{0}{0}{\rm T_{A}}$ as a sum of four matrices
\begin{align}\label{rhoviaABC}
\left(\rho\suptiny{0}{0}{\rm FS}\subtiny{0}{0}{ABC}\right)\suptiny{0}{0}{\rm T_{A}}=A+B+C+D,
\end{align}
where
\begin{align}
A&=(1-\lambda^2)\ket{000}\!\!\bra{000}\,+\,\tfrac{(1-\lambda^2)}{3}\tfrac{\lambda}{\sqrt{1-\lambda^2}}
    \bigl(\ket{000}\!\!\bra{\psi}+\ket{\psi}\!\!\bra{000}\bigr)\,+\,\tfrac{(1-\lambda^2)}{3}\tfrac{\lambda^2}{1-\lambda^2}\ket{\psi}\!\!\bra{\psi},
    \label{A}\\[0.5mm]
B&=\tfrac{(1-\lambda^2)}{3}\!\sum_{m=1}\suptiny{0}{0}\infty\! \lambda^{2m}
    \bigl(\ket{mm0}\!\!\bra{mm0}\,+\,\ket{m0m}\!\!\bra{m0m}\bigr),\label{B}\\[0.5mm]
C&=\tfrac{(1-\lambda^2)}{3}\sum_{m=1}\suptiny{0}{0}\infty\lambda^{m}
    \bigl(\ket{0m0}\!\!\bra{m00}\,+\,\ket{00m}\!\!\bra{m00}\,+\,\ket{m00}\!\!\bra{0m0}\,+\,\ket{m00}\!\!\bra{00m}\bigr),\label{C}\\[0.5mm]
D&=\tfrac{(1-\lambda^2)}{3}\!\sum_{m\ne n=1}\suptiny{0}{0}\infty\!\lambda^{m+n}
    \bigl(\ket{nm0}\!\!\bra{mn0}\,+\,\ket{n0m}\!\!\bra{m0n}\bigr),\label{D}
\end{align}
with $\ket{\psi}=\frac{\sqrt{1-\lambda^2}}{\lambda}\sum_{n=1}\suptiny{0}{0}\infty\lambda^{n}\ket{0nn}$ being a normalized vector. 
One can observe that $\left(\rho\suptiny{0}{0}{\rm FS}\subtiny{0}{0}{ABC}\right)\suptiny{0}{0}{\rm T_{A}}$ is block-diagonal with a corresponding splitting of the entire Hilbert space into the direct sum 
\begin{align}\label{Hsum}
\mathscr{H}=\mathscr{H}^{(A)}\oplus\mathscr{H}^{(B)}\oplus\mathscr{H}^{(C)}\oplus\mathscr{H}^{(D)}\oplus\mathscr{H}^{(O)},    
\end{align}
where, $\mathscr{H}^{(A)}$, $\mathscr{H}^{(B)}$, $\mathscr{H}^{(C)}$, and $\mathscr{H}^{(D)}$ are orthogonal invariant subspaces of the matrix 
$\left(\rho\suptiny{0}{0}{\rm FS}\subtiny{0}{0}{ABC}\right)\suptiny{0}{0}{\rm T_{A}}$ spanned by the vectors $\mathcal{A}=\{\ket{0mm},\,m\geq0\}$, $\mathcal{B}=\{\ket{mm0},\,\ket{m0m},\,m>0\}$, $\mathcal{C}=\{\ket{00m},\,\ket{0m0},\,\ket{m00},\,m>0\}$, and 
$\mathcal{D}=\{\ket{nm0},\,\ket{mn0},\,\ket{n0m},\,\ket{m0n},\,m\ne n,\,m,n>0\}$, while $\mathscr{H}^{(O)}$ is the null space spanned by all 
remaining three-mode Fock states including, e.g., the states $\{\ket{mmm},m>0\}$. 
This allows us to reduce our investigation to the analysis of the eigenvalues of the matrices corresponding to the individual blocks, $A$, $B$, $C$, and $D$.\\[-2mm]

The matrix $A$ has only one non-zero two-dimensional block, which corresponds to the orthonormal vectors 
$\{\ket{000},\,\ket{\psi}\}$, and which has two non-negative eigenvalues. Similarly, the matrix $B$ is already diagonal and the matrix $D$ splits into four-dimensional blocks each corresponding to the basis vectors $\{\ket{nm0},\,\ket{mn0},\,\ket{n0m},\,\ket{m0n}\}$ with fixed $m$ and $n$.\\[-2mm] 

Finally, for the task considered here the most important matrix is $C$. We see that this matrix consists of $3\times 3$ blocks, where each block corresponds to the set of vectors $\{\ket{00m},\ket{0m0},\ket{m00}\}, $where $m>0$ is fixed,  
and is of the form
\vspace*{-2mm}
\begin{align}
\tfrac{(1-\lambda^2)}{3}\left(\begin{array}{ccc}
        0 & 0 & \lambda^m \\
        0 & 0 & \lambda^m \\
        \lambda^m & \lambda^m & 0
    \end{array}
    \right).
\end{align}
The latter matrix possesses two eigenvalues $\mu_{m}^{\pm}=\pm\frac{\sqrt 2}{3}(1-\lambda^2)\lambda^{m}$ and one zero eigenvalue. A normalized eigenvector corresponding to the eigenvalue $\mu_{m}^{-}$ reads
\begin{align}\label{mum}
\ket{\mu_{m}^{-}}=\tfrac{1}{2}\left(\ket{00m}+\ket{0m0}-\sqrt{2}\ket{m00}\right),\quad m>0.
\end{align}
Since $\mu_{m}^{-}=-\tfrac{\sqrt 2}{3}(1-\lambda^2)\lambda^{m}<0$ for all $1>\lambda>0$, the density matrix $\rho\suptiny{0}{0}{\rm FS}\subtiny{0}{0}{ABC}$ is entangled across the partition 
$A|BC$ for all $r>0$, and due to its symmetry with respect to the exchange of the mode labels, the state is entangled with respect to all three bipartite splits. Consequently, 
the density matrix $\rho\suptiny{0}{0}{\rm FS}\subtiny{0}{0}{ABC}$ is fully inseparable for all $r>0$, as we set out to prove.

\subsection{GME activatability of biseparable non-Gaussian states}\label{SM:1.2}

For detecting GME activatability we turn to the $k$-separability criterion proposed in Ref.~\cite{GabrielHiesmayrHuber2010}: Every $k$-separable $N$-partite state $\rho$ satisfies
\begin{align}\label{cr:Gabriel}
    \sqrt{\bra{\phi}\rho^{\otimes 2}P_{\mathrm{tot}}\ket{\phi}}
    \leq
    \sum_{\{M\}}\Bigl(\prod_{i=1}^k\bra{\phi}P^\dagger_{M(i)}\rho^{\otimes 2}P_{M(i)}\ket{\phi}\Bigr)\suptiny{0}{0}{\tfrac{1}{2k}},
\end{align}
for every fully separable $2N$-partite state 
$\ket{\phi}=\bigotimes_{i=1}^{2N}\ket{\phi\subtiny{0}{0}{i}}$, where $P_{M(i)}$ are permutation operators that exchange the two copies of all subsystems contained in the $i$-th subset of the partition $M$, $P_{\mathrm{tot}}$ is an operator exchanging the two copies entirely, and the sum runs over all possible partitions $M$ of the considered system into $k$ subsystems.\\[-3mm]

We now employ this criterion for $k=2$ to check if two copies of $\rho\suptiny{0}{0}{\rm FS}\subtiny{0}{0}{ABC}$ from Eq.~(\ref{FS}) are GME.
In this case the state $\rho$ in Ineq.~(\ref{cr:Gabriel}) 
is
\begin{align}
    \rho=\rho\suptiny{0}{0}{\rm FS}\subtiny{0}{0}{A_{1}B_{1}C_{1}}\otimes\rho\suptiny{0}{0}{\rm FS}\subtiny{0}{0}{A_{2}B_{2}C_{2}},
    \label{eq:two copies nonGaus GME crit}
\end{align}
and we choose $\ket{\phi}$ to be the fully separable state 
\begin{align}
    \ket{\phi}  &=\ket{n00}\subtiny{0}{0}{A_{1}B_{1}C_{1}}\ket{0n0}\subtiny{0}{0}{A_{2}B_{2}C_{2}}\ket{n11}\subtiny{0}{0}{A' _{1}B'_{1}C'_{1}}\ket{1n1}\subtiny{0}{0}{A' _{2}B'_{2}C'_{2}}\,.
    \label{eq:fsep}
\end{align}
For this choice, the left-hand side of Ineq.~(\ref{cr:Gabriel}) takes the form
\begin{align}
    \sqrt{\bra{\phi}\rho^{\otimes 2}P_{\mathrm{tot}}\ket{\phi}}
    &=
    \sqrt{\bra{n000n0n111n1}\rho^{\otimes 2}\ket{n111n1n000n0}}=\,|\bra{n000n0}\rho\ket{n111n1}|=\\[1mm]
    &\ \ =\,|\bra{n00}\rho\suptiny{0}{0}{\rm FS}\subtiny{0}{0}{ABC}\ket{n11}|\times
    |\bra{0n0}\rho\suptiny{0}{0}{\rm FS}\subtiny{0}{0}{ABC}\ket{1n1}|
    \,=\,\tfrac{1}{9}(1-\lambda^{2})^{2}\lambda^{2}\,,\nonumber
\end{align}
where $P_{\mathrm{tot}}$ exchanges the primed and unprimed subsystems with each other, and in going from the second to the third line, we have used Eq.~(\ref{eq:two copies nonGaus GME crit}).\\[-3mm]

The right-hand side of Ineq.~(\ref{cr:Gabriel}) is a sum of three terms corresponding to the three bipartitions $A_{1}A_{2}|B_{1}B_{2}C_{1}C_{2}$, $A_{1}A_{2}B_{1}B_{2}|C_{1}C_{2}$, and $A_{1}A_{2}C_{1}C_{2}|B_{1}B_{2}$. 
Each of these terms is a square root, and the arguments of these square roots are products of diagonal density-matrix elements. 
Specifically, for the bipartition $A_{1}A_{2}|B_{1}B_{2}C_{1}C_{2}$ there are two factors, one obtained by exchanging the subsystem $A_{1}A_{2}$ with $A'_{1}A'_{2}$, the other by exchanging $B_{1}B_{2}C_{1}C_{2}$ with $B'_{1}B'_{2}C'_{1}C'_{2}$, such that we have
\begin{align}
    & \bra{n00\,1n0\,n11\,0n1}\rho^{\otimes 2}\ket{n00\,1n0\,n11\,0n1}\bra{n11\,0n1\,n00\,1n0}\rho^{\otimes 2}\ket{n11\,0n1\,n00\,1n0}\nonumber\\[1mm]
 &\ \ \ =\,|\bra{n00\,1n0}\rho\ket{n00\,1n0}|^{2}\times
 |\bra{n11\,0n1}\rho\ket{n11\,0n1}|^{2}
 \\[1mm]
&\ \ \ =\,|\bra{n00}\rho\suptiny{0}{0}{\rm FS}\subtiny{0}{0}{ABC}\ket{n00}|^{2}\,\times
|\bra{1n0}\rho\suptiny{0}{0}{\rm FS}\subtiny{0}{0}{ABC}\ket{1n0}|^{2}|\bra{n11}\rho\suptiny{0}{0}{\rm FS}\subtiny{0}{0}{ABC}\ket{n11}|^{2}
\times
|\bra{0n1}\rho\suptiny{0}{0}{\rm FS}\subtiny{0}{0}{ABC}\ket{0n1}|^{2}
 \,=\,0\,,\nonumber
\end{align}
which vanishes because the matrix elements $|\bra{1n0}\rho\suptiny{0}{0}{\rm FS}\subtiny{0}{0}{ABC}\ket{1n0}|=0$ and $|\bra{0n1}\rho\suptiny{0}{0}{\rm FS}\subtiny{0}{0}{ABC}\ket{0n1}|=0$.
Similarly, the arguments of the square roots for the other two bipartitions evaluate to
\begin{align}
    &
    \bra{n10\,1n0\,n01\,0n1}\rho^{\otimes 2}\ket{n10\,1n0\,n01\,0n1}\bra{n01\,0n1\,n10\,1n0}\rho^{\otimes 2}\ket{n01\,0n1\,n10\,1n0}
    \nonumber\\[1mm]
 &\ \ \ =\,|\bra{n01\,0n1}\rho\ket{n01\,0n1}|^{2}\times
 |\bra{n10\,1n0}\rho\ket{n10\,1n0}|^{2}
 \\[1mm]
&\ \ \ =\,|\bra{n01}\rho\suptiny{0}{0}{\rm FS}\subtiny{0}{0}{ABC}\ket{n01}|^{2}\,\times
|\bra{0n1}\rho\suptiny{0}{0}{\rm FS}\subtiny{0}{0}{ABC}\ket{0n1}|^{2}
|\bra{n10}\rho\suptiny{0}{0}{\rm FS}\subtiny{0}{0}{ABC}\ket{n10}|^{2}
\times
|\bra{1n0}\rho\suptiny{0}{0}{\rm FS}\subtiny{0}{0}{ABC}\ket{1n0}|^{2}
 \,=\,0\,.\nonumber
\end{align}
and
\begin{align}
    &
    \bra{n10\,0n0\,n01\,1n1}\rho^{\otimes 2}\ket{n10\,0n0\,n01\,1n1}\bra{n01\,1n1\,n10\,0n0}\rho^{\otimes 2}\ket{n01\,1n1\,n10\,0n0}
    \nonumber\\[1mm]
 &\ \ \ =\,|\bra{n10\,0n0}\rho\ket{n10\,0n0}|^{2}\times
 |\bra{n01\,1n1}\rho\ket{n01\,1n1}|^{2}
 \\[1mm]
&\ \ \ =\,|\bra{n10}\rho\suptiny{0}{0}{\rm FS}\subtiny{0}{0}{ABC}\ket{n10}|^{2}\,\times
|\bra{0n0}\rho\suptiny{0}{0}{\rm FS}\subtiny{0}{0}{ABC}\ket{0n0}|^{2}
|\bra{n01}\rho\suptiny{0}{0}{\rm FS}\subtiny{0}{0}{ABC}\ket{n01}|^{2}
\times
|\bra{1n1}\rho\suptiny{0}{0}{\rm FS}\subtiny{0}{0}{ABC}\ket{1n1}|^{2}
 \,=\,0\,.\nonumber
\end{align}
Since the right-hand side of Ineq.~(\ref{cr:Gabriel}) vanishes, and the left-hand side is larger than zero for all $r\neq 0$, we see that all FIB states in this family are GME activatable.


\section{GME activation for non-identical Gaussian states}\label{SM:2}

The CM biseparability criterion~\cite{HyllusEisert2006} states that for any biseparable state with CM $\gamma\subtiny{0}{0}{\mathrm{BS}}$ there exist block-diagonal CMs $\gamma\subtiny{0}{0}{M(i)}$ corresponding to the partition $M(i)$ along with a probability distribution $\{p_i\}$ such that 
\begin{align}\label{bisepCM}
\gamma\subtiny{0}{0}{\mathrm{BS}}-\sum_i \,p_i\,\gamma\subtiny{0}{0}{M(i)}\,\geq\,0\,.
\end{align}
In the main text, we have shown that the CM biseparability criterion~(\ref{bisepCM}) cannot detect GME activation for $k$ identical copies since the CM of $\rho^{\otimes k}$ automatically satisfies the criterion if the criterion is satisfied by the CM of $\rho$. 
This is the case independently of the Gaussian or non-Gaussian character of the state.\\[-3mm]

However, as we will demonstrate here, the CM biseparability criterion can be used to detect GME activation for (certain) non-identical pairs of states.
In the case of the proof for $k$ identical copies in the main text, we considered that if for a given CM $\gamma\subtiny{0}{0}{\mathrm{BS}}$ the ensemble $\{(p_i,\gamma\subtiny{0}{0}{M(i)})\}_{i}$ is such that $\Delta\gamma:=\gamma\subtiny{0}{0}{\mathrm{BS}}-\sum_i \,p_i\,\gamma\subtiny{0}{0}{M(i)}\geq0$, then the ensemble $\{(p_i,\bigoplus_{n=1}^{k}\gamma\subtiny{0}{0}{M(i)}\suptiny{0}{0}{(n)})\}_{i}$ with $\gamma\subtiny{0}{0}{M(i)}\suptiny{0}{0}{(n)}=\gamma\subtiny{0}{0}{M(i)}\,\forall\,n$ satisfies
\begin{align}\label{eq:result}
    \bigoplus_{n=1}^{k}\gamma\subtiny{0}{0}{\mathrm{BS}}
    -
    \sum_i \,p_i\,\bigoplus_{n=1}^{k}\gamma\subtiny{0}{0}{M(i)}\suptiny{0}{0}{(n)}
    \,=\,
    \bigoplus_{n=1}^{k}\bigl(\gamma\subtiny{0}{0}{\mathrm{BS}}-
    \sum_i \,p_i\,\gamma\subtiny{0}{0}{M(i)}\bigr)\,\geq\,0
    \,,
\end{align}
since the left-hand side is block-diagonal and each block is identical to a positive semi-definite matrix $\Delta\gamma\geq0$.
Here, the equality holds under the condition that the two CMs in question admit decompositions into convex sums (with each term in the sum a CM that is block-diagonal with respect to one of the bipartitions) with the same probability distribution $\{p_i\}_{i}$. 
That is, for two CMs $\gamma$ and $\tilde{\gamma}$ that admit decompositions $\{(p_i,\gamma\subtiny{0}{0}{M(i)})\}_{i}$ and $\{(p_i,\tilde{\gamma}\subtiny{0}{0}{M(i)})\}_{i}$ such that
\begin{subequations}
\begin{align}
    \gamma-\sum_i \,p_i\,\gamma\subtiny{0}{0}{M(i)} &\geq\,0\,,\\[1mm]
    \tilde{\gamma}-\sum_i \,p_i\,\tilde{\gamma}\subtiny{0}{0}{M(i)} &\geq\,0\,,
\end{align}
\end{subequations}
the CM $\gamma\oplus\tilde{\gamma}$ of the joint state still satisfies
\begin{align}
    \gamma\oplus\tilde{\gamma}-\sum_i \,p_i\,\gamma\subtiny{0}{0}{M(i)}\oplus\tilde{\gamma}\subtiny{0}{0}{M(i)}
    \,=\,\bigl(\gamma-\sum_i \,p_i\,\gamma\subtiny{0}{0}{M(i)}\bigr)
    \oplus \bigl(\tilde{\gamma}-\sum_i \,p_i\,\tilde{\gamma}\subtiny{0}{0}{M(i)} \bigr)\,
    \geq\,0\,.
\end{align}
This line of reasoning generally no longer goes through if the two CMs do not admit decompositions with the same probability distributions $\{p_i\}_i$. In the next part, we will show a simple example of two different CMs satisfying Ineq.~(\ref{bisepCM}) that together create a CM of tripartite GME state, where each party is composed of two modes. 

\begin{figure}[t]
    \centering
    \includegraphics[scale=0.75]{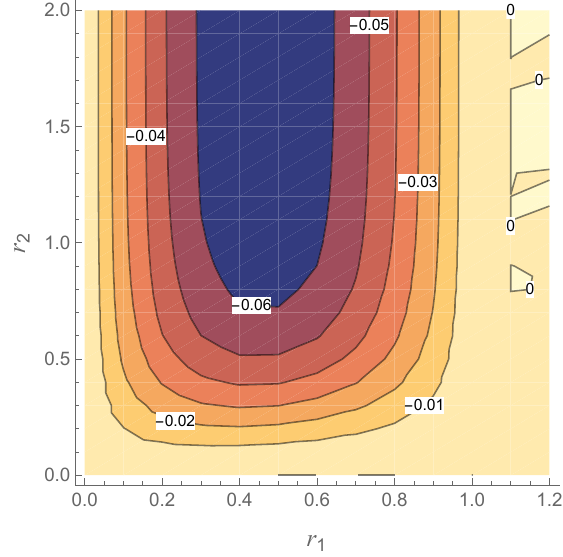}
    \caption{The quantity $\mathrm{Tr}(\mathcal{W}\gamma\subtiny{0}{0}{AaBbCc}(r_1, r_2))-1$ for ranges of parameters $r_1$ and $r_2$ such that original states are fully inseparable and biseparable. Negative values indicate a violation of Ineq.~(\ref{bisepCM}) and thus GME of the compounded state $\rho\subtiny{0}{0}{ABC}\suptiny{0}{0}{\rm FS}\otimes\rho\subtiny{0}{0}{abc}\suptiny{0}{0}{\rm PS}$.}
    \label{fig:2diffCopies}
\end{figure}

Let us take the CM of the biseparable state, Eq.~\eqref{FS}, for $n=0$,
\begin{align}\label{testCM}
    \gamma\subtiny{0}{0}{ABC}(r)
    =\tfrac{1}{3}\bigl(
    \gamma\subtiny{0}{0}{AB}\suptiny{0}{0}{\mathrm{TMSV}} \oplus \mathds{1}\subtiny{0}{0}{C}
    +\gamma\subtiny{0}{0}{BC}\suptiny{0}{0}{\mathrm{TMSV}} \oplus \mathds{1}\subtiny{0}{0}{A}
    +\gamma\subtiny{0}{0}{AC}\suptiny{0}{0}{\mathrm{TMSV}} \oplus \mathds{1}\subtiny{0}{0}{B}\bigr)
\end{align}
and the CM of the biseparable state $\rho\suptiny{0}{0}{\rm PS}\subtiny{0}{0}{abc} = \frac{1}{2} (\rho\subtiny{0}{0}{ab}\suptiny{0}{0}{\rm TMSV} \otimes \rho\subtiny{0}{0}{c}\suptiny{0}{0}{sq} + \rho\subtiny{0}{0}{bc}\suptiny{0}{0}{\rm TMSV} \otimes \rho\subtiny{0}{0}{a}\suptiny{0}{0}{sq})$ $\gamma\subtiny{0}{0}{abc}$ given by
\begin{align}\label{partsymCM}
    \gamma\subtiny{0}{0}{abc}(r)
    =\tfrac{1}{2}\bigl(
    \gamma\subtiny{0}{0}{ab}\suptiny{0}{0}{\mathrm{TMSV}} \oplus \gamma\subtiny{0}{0}{c}\suptiny{0}{0}{\mathrm{sq}}
    +\gamma\subtiny{0}{0}{bc}\suptiny{0}{0}{\mathrm{TMSV}} \oplus \gamma\subtiny{0}{0}{a}\suptiny{0}{0}{\mathrm{sq}}\bigr).
\end{align}
Here
\begin{align}
    \gamma\suptiny{0}{0}{\mathrm{TMSV}}=\left(\begin{array}{cc}
       \cosh(2r)\mathds{1}  & \sinh(2r)Z \\
        \sinh(2r)Z & \cosh(2r)\mathds{1}
    \end{array}
    \right)
\end{align}
is the CM of a TMSV state, $Z=\operatorname{diag}\{1,-1\}$ is the usual third Pauli matrix, $\mathds{1}$ is the CM of the single-mode vacuum state, and $\gamma\suptiny{0}{0}{\mathrm{sq}}=\mbox{diag}(e^{2r},e^{-2r})$ is CM of single-mode squeezed vacuum state $\rho\suptiny{0}{0}{sq}$. The CM $\gamma\subtiny{0}{0}{AaBbCc}(r_1, r_2) = \gamma\subtiny{0}{0}{ABC}(r_1)\oplus\gamma\subtiny{0}{0}{abc}(r_2)$ is no longer of the form of the decomposition $\sum_i p_i\gamma\subtiny{0}{0}{M(i)}\oplus\tilde{\gamma}\subtiny{0}{0}{M(i)}$ with the same probabilities $\{p_i\}_i$. To test the GME of the product state $\rho\subtiny{0}{0}{ABC}\suptiny{0}{0}{\rm FS} \otimes \rho\subtiny{0}{0}{abc}\suptiny{0}{0}{\rm PS}$, we applied a semi-definite program~\cite{HyllusEisert2006} to find an optimal GME witness matrix $\mathcal{W}$ for CM $\gamma\subtiny{0}{0}{AaBbCc}(r_1, r_2)$. This is a real symmetric positive semi-definite matrix for which $\mathrm{Tr}(\mathcal{W}\gamma\subtiny{0}{0}{\rm BS})-1\geq0$ for all biseparable CMs and $\mathrm{Tr}(\mathcal{W}\gamma)-1<0$ for at least one CM~\cite{HyllusEisert2006}. 
We searched through ranges $r_1\in (0,1.24)$ and $r_2\in (0,2)$, where the CMs, Eq.~\eqref{testCM} and Eq.~\eqref{partsymCM}, exhibit full inseparability, and we detected GME of the compounded state for almost all values of $r_1$ and $r_2$ (see Fig.~\ref{fig:2diffCopies} for more details).\\
Thus, although the criterion (\ref{bisepCM}) cannot be used to detect GME activation in the case of multiple identical copies of a biseparable state, it can, in principle, detect GME of the product of different FIB states. 

\section{GME detection for Gaussian states satisfying the CM biseparability criterion}\label{SM:3}

In this section, we focus on a specific one-parameter family of Gaussian states $\rho\suptiny{0}{0}{\rm G}\subtiny{0}{0}{ABC}(r)$ with a vanishing vector of first moments and CM by Eq.~(\ref{testCM}).
In Sec.~\ref{SM:3.1}, we study the range of the parameter $r$ for which the state is fully inseparable. 
In Sec.~\ref{SM:3.2}, we then present a GME witness that is able to detect a range of $r$ for which the states $\rho\suptiny{0}{0}{\rm G}\subtiny{0}{0}{ABC}(r)$ are certainly GME. 
We describe the calculation of the required density-matrix elements of $\rho\suptiny{0}{0}{\rm G}\subtiny{0}{0}{ABC}(r)$ in Sec.~\ref{SM:3.3}. 
Finally, in Sec.~\ref{SM:3.4}, we use these density-matrix elements to construct a three-qubit state and analyze its entanglement structure.\\[-2.5mm]

Before we proceed, let us make a brief remark regarding the parameter $r$. The CM $\gamma\subtiny{0}{0}{ABC}$ in Eq.~(\ref{testCM}) is a convex combination of CMs corresponding to product states of TMSV states and vacuum states for the third mode, with each term in the convex combination corresponding to a different labeling of the modes.
For each individual term, the parameter $r$ represents a (two-mode) squeezing parameter that directly relates to the bipartite entanglement between the corresponding pair of modes. 
However, as we see here, the convex combination of CMs is \emph{not} equivalent to a convex combination of the corresponding density matrices.
As such, the parameter $r$ can no longer be interpreted as a squeezing parameter in the usual sense of parameterizing a unitary (two-mode squeezing) transformation that monotonously increases the entanglement between two modes that are initially in a pure product state (the vacuum). 
Indeed, here the purity $P(\rho\suptiny{0}{0}{\rm G}\subtiny{0}{0}{ABC})=1/\sqrt{\det(\gamma\subtiny{0}{0}{ABC})}$ of the three-mode state we consider decreases with increasing $r$. 
Specifically, the determinant of the CM is given by
\vspace*{-1.5mm}
\begin{align}
    \det(\gamma\subtiny{0}{0}{ABC}) &=\,
    \bigl(5 + 4 \cosh(2r)\bigr)\Bigl(\tfrac{7 + 8 \cosh(2r) + 3 \cosh(4r)}{54}\Bigr)^{2}\,.
\end{align}
At the same time, we note that for $r=0$ the CM reduces to $\gamma\subtiny{0}{0}{ABC}(r=0)=\mathds{1}\subtiny{0}{0}{A}\oplus \mathds{1}\subtiny{0}{0}{B}\oplus \mathds{1}\subtiny{0}{0}{C}$ and $\rho\suptiny{0}{0}{\rm G}\subtiny{0}{0}{ABC}(r=0)$ is hence the fully separable vacuum state, $\ket{0}\subtiny{0}{0}{A}\ket{0}\subtiny{0}{0}{B}\ket{0}\subtiny{0}{0}{C}$. 
Already from these observations, it is thus expected that any non-trivial bipartite and multipartite entanglement will appear for $r>0$ but only up to a certain value of $r$, at which the increasing mixedness of the three-mode state and of the single-mode reduced states suppresses any quantum correlations between the modes. 
In the next section, we will quantify this intuition.


\subsection{Range of full inseparbility}\label{SM:3.1}

Here we determine the range of the parameter $r$ for which the Gaussian state $\rho\suptiny{0}{0}{\rm G}\subtiny{0}{0}{ABC}(r)$ described by the CM $\gamma\subtiny{0}{0}{ABC}$ from Eq.~(\ref{testCM}) is fully inseparable (i.e., FIB or GME). 
Generally, a tripartite state is fully separable if it is separable with respect to all bipartitions. 
Here, given the symmetry of the state concerning the exchange of the mode labels, this means we just have to check for separability with respect to any fixed bipartition. 
Without loss of generality, we consider the bipartition $AB|C$ and apply the PPT criterion, which provides a necessary and sufficient criterion for separability of $1$ vs. $N$-mode Gaussian states~\cite{WernerWolf2001}.\\[-3mm]

On the level of the CM, the partial transposition can be represented as a flip of the momentum quadrature of the respective single mode (here, mode $C$),
    $\gamma\subtiny{0}{0}{ABC}\mapsto \tilde{\gamma}\subtiny{0}{0}{ABC}=\tilde{T}\subtiny{0}{0}{C}\gamma\subtiny{0}{0}{ABC}\tilde{T}\subtiny{0}{0}{C}$,
where $\tilde{T}\subtiny{0}{0}{C}=\mathds{1}\subtiny{0}{0}{AB}\oplus Z\subtiny{0}{0}{C}$ and $Z=\operatorname{diag}\{1,-1\}$ is the usual third Pauli matrix. 
Then, the corresponding Gaussian state is entangled with respect to the bipartition $AB|C$ if the smallest symplectic eigenvalue $\tilde{\nu}_{-}$ of $\tilde{\gamma}\subtiny{0}{0}{ABC}$ is smaller than $1$. The quantity $\tilde{\nu}_{-}$ can be calculated as the smallest eigenvalue of $|i\,\Omega\,\tilde{\gamma}\subtiny{0}{0}{ABC}|$ with the symplectic form
\begin{align}\label{symplectic form}
\Omega=\bigoplus^{3}_{i=1}
\begin{pmatrix}
0&1\\
-1&0
\end{pmatrix}.
\end{align}
As a function of $r$, we find that the smallest symplectic eigenvalue of the `partially transposed' CM is given by
\begin{align}
    &\tilde{\nu}_{-} \,=\,
    \tfrac{1}{6}\Bigl(
    \,9 + 16 \cosh(2r) + 11 \cosh(4r) - \sqrt{2\sinh^{\hspace*{-1pt}2}(2r)
    \bigl[199 + 256 \cosh(2r) + 121 \cosh(4r)\bigr]
    }\
    \Bigr)\suptiny{0}{0}{1/2}.\nonumber
\end{align}
The condition $\tilde{\nu}_{-}(r=r_{1})\,=\,1$ then determines the value $r=r_1$ at which the state becomes separable with respect to the chosen bipartition, and hence separable with respect to all bipartitions. This condition can then be seen to be equivalent to the condition $47 + 28 \cosh(2r)-3 \cosh(4r) = \,0$, which is solved by
\vspace*{-1.5mm}
\begin{align}
    r_1&=\,\tfrac{1}{2}\operatorname{arcosh}\bigl(\tfrac{7 + 2 \sqrt{31}}{3}\bigr)\approx\,1.24275\,.
    \label{eq:r1}
\end{align}
\subsection{GME Witness inequality}\label{SM:3.2}

We now present a (non-linear) GME witness inequality that is a generalization of a witness that appeared as Eq.~(A4) in Ref.~\cite{FriisHuberFuentesBruschi2012}, using techniques similar to the witnesses derived in Ref.~\cite{GabrielHiesmayrHuber2010} and Ref.~\cite{HuberMintertGabrielHiesmayr2010}. 
All tripartite biseparable states satisfy
\begin{align}
\begin{split}\label{3modewitness}
    &
    |\bra{000}\rho\suptiny{0}{0}{\rm BS}\subtiny{0}{0}{ABC}\ket{011}|
    +
    |\bra{000}\rho\suptiny{0}{0}{\rm BS}\subtiny{0}{0}{ABC}\ket{101}|
    +
    |\bra{000}\rho\suptiny{0}{0}{\rm BS}\subtiny{0}{0}{ABC}\ket{110}|\,\\[1.5mm]
    &
    \leq\sqrt{\bra{000}\rho\suptiny{0}{0}{\rm BS}\subtiny{0}{0}{ABC}\ket{000}}\sqrt{
    \bra{011}\rho\suptiny{0}{0}{\rm BS}\subtiny{0}{0}{ABC}\ket{011}
    +
    \bra{101}\rho\suptiny{0}{0}{\rm BS}\subtiny{0}{0}{ABC}\ket{101}
    +
    \bra{110}\rho\suptiny{0}{0}{\rm BS}\subtiny{0}{0}{ABC}\ket{110}
    }\\[1.5mm]
    &
    \qquad+\sqrt{\bra{001}\rho\suptiny{0}{0}{\rm BS}\subtiny{0}{0}{ABC}\ket{001}\,
    \bra{010}\rho\suptiny{0}{0}{\rm BS}\subtiny{0}{0}{ABC}\ket{010}}+\sqrt{\bra{001}\rho\suptiny{0}{0}{\rm BS}\subtiny{0}{0}{ABC}\ket{001}\,
    \bra{100}\rho\suptiny{0}{0}{\rm BS}\subtiny{0}{0}{ABC}\ket{100}}\\[1.5mm]
    &
    \qquad+\sqrt{\bra{010}\rho\suptiny{0}{0}{\rm BS}\subtiny{0}{0}{ABC}\ket{010}\,
    \bra{100}\rho\suptiny{0}{0}{\rm BS}\subtiny{0}{0}{ABC}\ket{100}}\,.
    \end{split}
\end{align}

\noindent\textbf{Proof}.\ 
To show that this inequality holds for all biseparable states, we first show that it holds for a product state $\rho$ for a fixed bipartition; without loss of generality, we choose the bipartition $A|BC$. 
From the symmetry of the inequality with respect to the exchange of the subsystems, it then follows that the inequality holds for product states for any bipartition. 
Finally, the validity for arbitrary convex mixtures of such states follows from the convexity of the absolute values on the left-hand side and from the concavity of the square roots on the right-hand side. 
More specifically, the left-hand side of Ineq.~(\ref{3modewitness}) is a sum of convex functions, and is hence itself a convex function $f(\rho)$ of the density operator $\rho$, while the right-hand side is a sum of concave functions, and hence itself a concave function $g(\rho)$: When replacing the (product) state $\rho$ with a convex mixture of (product) states $\rho_i$, that is, for $\rho\rightarrow\sum_{i}p_{i}\rho_{i}$, we can upper bound the function $f$ evaluated on the mixture of states $\rho_{i}$ by the convex mixture of the function evaluated on the different $\rho_i$, $f(\sum_{i}p_{i}\rho_{i})\leq \sum_{i}p_{i}f(\rho_{i})$. We then apply the original inequality for product states, $f(\rho_{i})\leq g_{j}(\rho_{i})$, and use the concavity of $g$, i.e., $\sum_{i}p_{i}g(\rho_{i})\leq g(\sum_{i}p_{i}\rho_{i})$, to show that Ineq.~(\ref{3modewitness}) holds for all biseparable states.\\[-2.5mm]

To see that the inequality holds for a product state for the bipartition $A|BC$, we set $\rho\suptiny{0}{0}{\rm BS}\subtiny{0}{0}{ABC}=\rho\subtiny{0}{0}{A}\otimes\rho\subtiny{0}{0}{BC}$, such that the left-hand side of Ineq.~(\ref{3modewitness}) becomes
\begin{align}
\begin{split}\label{3modewitnessproof}
    &
    |\bra{000}\rho\suptiny{0}{0}{\rm BS}\subtiny{0}{0}{ABC}\ket{011}|
    +
    |\bra{000}\rho\suptiny{0}{0}{\rm BS}\subtiny{0}{0}{ABC}\ket{101}|
    +
    |\bra{000}\rho\suptiny{0}{0}{\rm BS}\subtiny{0}{0}{ABC}\ket{110}|\,\\[1.5mm]
    &
    =\,\bra{0}\rho\subtiny{0}{0}{A}\ket{0}\times |\bra{00}\rho\subtiny{0}{0}{BC}\ket{11}|\,+\,|\bra{0}\rho\subtiny{0}{0}{A}\ket{1}|\times
    |\bra{00}\rho\subtiny{0}{0}{BC}\ket{01}|\,+\,|\bra{0}\rho\subtiny{0}{0}{A}\ket{1}|\times
    |\bra{00}\rho\subtiny{0}{0}{BC}\ket{10}|\,.
\end{split}
\end{align}
We then use the spectral decomposition of any state $\rho=\sum_{i}p_{i}\ket{\psi_{i}}\!\!\bra{\psi_{i}}$ along with the Cauchy-Schwarz inequality $|\vec{x}\cdot\vec{y}\,|\leq|\vec{x}\,|\cdot|\vec{y}\,|$ to write
\begin{align}
\begin{split}
    &|\bra{m}\rho\ket{n}|    \ =\ \Bigl|\sum\limits_{i}\sqrt{p_{i}}\scpr{m}{\psi_{i}}\,\sqrt{p_{i}}\scpr{\psi_{i}}{n}\Bigr|\\[0.0mm]
    &\ \leq\,
    \,\sqrt{\sum\limits_{i}p_{i}|\scpr{m}{\psi_{i}}|^{2}}\,
    \sqrt{\sum\limits_{j}p_{j}|\scpr{n}{\psi_{j}}|^{2}}=\,\sqrt{\bra{m}\rho\ket{m}\,\bra{n}\rho\ket{n}}\,.
\end{split}
\end{align}
With this, the terms on the right-hand side of Eq.~(\ref{3modewitnessproof}) can be bounded according to
\begin{align}
\begin{split}\label{3modewitnessproof2}
    &
    |\bra{000}\rho\suptiny{0}{0}{\rm BS}\subtiny{0}{0}{ABC}\ket{011}|
    +
    |\bra{000}\rho\suptiny{0}{0}{\rm BS}\subtiny{0}{0}{ABC}\ket{101}|
    +
    |\bra{000}\rho\suptiny{0}{0}{\rm BS}\subtiny{0}{0}{ABC}\ket{110}|\,\\[1.5mm]
    &
    \leq\,\bra{0}\rho\subtiny{0}{0}{A}\ket{0}\times
    \sqrt{\bra{00}\rho\subtiny{0}{0}{BC}\ket{00}\,
    \bra{11}\rho\subtiny{0}{0}{BC}\ket{11}}\,\\[1.5mm]
    &
    \ \ \ +\,
    \sqrt{\bra{0}\rho\subtiny{0}{0}{A}\ket{0}\,
        \bra{1}\rho\subtiny{0}{0}{BC}\ket{1}}
    \sqrt{\bra{00}\rho\subtiny{0}{0}{A}\ket{00}\,
    \bra{01}\rho\subtiny{0}{0}{BC}\ket{01}}\\[1.5mm]
    &
    \ \ \ +\,
    \sqrt{\bra{0}\rho\subtiny{0}{0}{A}\ket{0}\,
        \bra{1}\rho\subtiny{0}{0}{BC}\ket{1}}
    \sqrt{\bra{00}\rho\subtiny{0}{0}{A}\ket{00}\,
    \bra{10}\rho\subtiny{0}{0}{BC}\ket{10}
    }\,.
\end{split}
\end{align}
Now, a simple comparison with the right-hand side of Ineq.~(\ref{3modewitness}) for $\rho\suptiny{0}{0}{\rm BS}\subtiny{0}{0}{ABC}=\rho\subtiny{0}{0}{A}\otimes\rho\subtiny{0}{0}{BC}$ shows that each of the terms on the right-hand side of Ineq.~(\ref{3modewitnessproof2}) is matched by an equal or larger term on the right-hand side of Ineq.~(\ref{3modewitness}), thus showing that the inequality holds.\hfill

Since the Gaussian three-mode state $\rho\suptiny{0}{0}{\rm G}\subtiny{0}{0}{ABC}$ that we consider is fully symmetric with respect to the exchange of any two modes, the witness inequality from Ineq.~(\ref{3modewitness}) takes the more compact form
\begin{align}
    \sqrt{3}
    |\bra{000}\rho\suptiny{0}{0}{\rm G}\subtiny{0}{0}{ABC}\ket{011}|
    &\leq
    \sqrt{\bra{000}\rho\suptiny{0}{0}{\rm G}\subtiny{0}{0}{ABC}\ket{000}
    \bra{011}\rho\suptiny{0}{0}{\rm G}\subtiny{0}{0}{ABC}\ket{011}
    }\,
    +\,\sqrt{3}\,\bra{001}\rho\suptiny{0}{0}{\rm G}\subtiny{0}{0}{ABC}\ket{001}\,.
    \label{ineq FS state SM}
\end{align}
\subsection{Reconstruction of density-matrix elements from the Wigner function}\label{SM:3.3}

To use the witness from Ineq.~(\ref{ineq FS state SM}), we need to calculate density-matrix elements of the Gaussian state $\rho\subtiny{0}{0}{ABC}\suptiny{0}{0}{G}$ from its CM and vector of first moments, with the latter trivially being zero.
This calculation can be done via the Wigner function since the Wigner function of a Gaussian state is given as
\begin{align}\label{WigGauss SM}
    W(\mathbf{r})=\frac{e^{-(\mathbf{r}-\mathbf{d})^\mathrm{T}\gamma^{-1}(\mathbf{r}-\mathbf{d})}}{\pi^N\sqrt{\mathrm{det}\gamma}},
\end{align}
where $\mathbf{d}$ is the vector of first moments with elements $d_i=\left\langle {r}_i\right\rangle=\mathrm{Tr}({\rho}\,{r}_i)$ and $\gamma$ is the CM corresponding to the given Gaussian state with components
\begin{align}\label{CM}
\gamma_{ij}=\left\langle {r}_i {r}_j+ {r}_j {r}_i\right\rangle-2\left\langle {r}_i\right\rangle\left\langle {r}_j\right\rangle.
\end{align}
The Wigner function $W(\mathbf{x},\mathbf{p})[\rho\subtiny{0}{0}{ABC}\suptiny{0}{0}{G}]$ can be obtained directly by substituting the CM from Eq.~(\ref{testCM}) into Eq.~(\ref{WigGauss SM}) with $\mathbf{d}=0$. 
With the Wigner function at hand, we can use a relation
\begin{align}
    \tr\bigl({\rho}\,{G}\bigr)
    &=\,(2\pi)^{N}\!\!
    \int \!\!\mathrm{d}^N\!\mathbf{x}\,\mathrm{d}^N\!\mathbf{p}\,
    W(\mathbf{x},\mathbf{p})[{\rho}]\,
    W(\mathbf{x},\mathbf{p})[{G}]\,,
\end{align}
and thus, we can
obtain the density-matrix elements $\bra{i\subtiny{0}{0}{A}j\subtiny{0}{0}{B}k\subtiny{0}{0}{C}}\rho\subtiny{0}{0}{ABC}\suptiny{0}{0}{G}\ket{i'\subtiny{0}{0}{A}j'\subtiny{0}{0}{B}k'\subtiny{0}{0}{C}}$ by calculating
\begin{align}
\begin{split}
    &\bra{i\subtiny{0}{0}{A}j\subtiny{0}{0}{B}k\subtiny{0}{0}{C}}\rho\subtiny{0}{0}{ABC}\suptiny{0}{0}{G}\ket{i'\subtiny{0}{0}{A}j'\subtiny{0}{0}{B}k'\subtiny{0}{0}{C}}
    =\tr\left(\rho\subtiny{0}{0}{ABC}\suptiny{0}{0}{G}
    \ket{i\subtiny{0}{0}{A}}\!\!\bra{i'\subtiny{0}{0}{A}}
    \otimes
    \ket{j\subtiny{0}{0}{B}}\!\!\bra{j'\subtiny{0}{0}{B}}
    \otimes
    \ket{k\subtiny{0}{0}{C}}\!\!\bra{k'\subtiny{0}{0}{C}}
    \right)\\[1mm]
    &
    \ \ \ =\,(2\pi)^{3}\!\!
    \int \!\!\mathrm{d}^{3}\!\mathbf{x}\,\mathrm{d}^{3}\!\mathbf{p}\,
    W(\mathbf{x},\mathbf{p})[\rho\subtiny{0}{0}{ABC}\suptiny{0}{0}{G}] W(\mathbf{x},\mathbf{p})\bigl[
    \ket{i\subtiny{0}{0}{A}}\!\!\bra{i'\subtiny{0}{0}{A}}
    \otimes
    \ket{j\subtiny{0}{0}{B}}\!\!\bra{j'\subtiny{0}{0}{B}}
    \otimes
    \ket{k\subtiny{0}{0}{C}}\!\!\bra{k'\subtiny{0}{0}{C}}
    \bigr]\,,
\end{split}
\end{align}
where $W(\mathbf{x},\mathbf{p})[M]$ is the Wigner function for the matrix element in the argument in square brackets. 
Here, the states $\ket{i}$, $\ket{j}$, $\ket{k}$ and $\ket{i'}$, $\ket{j'}$, $\ket{k'}$ are single-mode Fock states.
Below, we provide expressions for the density-matrix elements in the subspace where each of the modes has at most one excitation, i.e., for $i,i',j,j',k,k'=\{0,1\}$. 
For the evaluation of the Wigner function, $W(\mathbf{x},\mathbf{p})\bigl[
    \ket{i\subtiny{0}{0}{A}}\!\!\bra{i'\subtiny{0}{0}{A}}
    \otimes
    \ket{j\subtiny{0}{0}{B}}\!\!\bra{j'\subtiny{0}{0}{B}}
    \otimes
    \ket{k\subtiny{0}{0}{C}}\!\!\bra{k'\subtiny{0}{0}{C}}
    \bigr]$ we further require the relation
\begin{align}
    \scpr{n}{x} &=\,\frac{(-1)^{n}e^{x^{2}/2}}{\sqrt{n!2^{n}\sqrt{\pi}}}
    \Bigl(\frac{d^n}{dx^n}e^{-x^2}\Bigr),
\end{align}
for the Fock-state wave functions. 
The calculation of the density-matrix elements then amounts to the evaluation of Gaussian integrals (nine for each matrix element, three each for the variables $\mathbf{x}$, $\mathbf{y}$, and $\mathbf{p}$) and algebraic simplification of the results. 
We start by defining the shorthand functions
\begin{subequations}
\begin{align}
    f(r)    &:=\,\tfrac{2}{\sqrt{5 + 4 \cosh(2r)}}\,, \qquad\rm and\\[1mm]
    g(r)    &:=\,\tfrac{9}{37 + 32 \cosh(2r) + 3 \cosh(4r)}\,.
\end{align}
\end{subequations}
We can then compactly write the matrix elements as
\begin{small}
\begin{subequations}\label{eq:DM matrix elements 3 qubits}
\begin{align}
     &\bra{000}\rho\suptiny{0}{0}{\rm G}\subtiny{0}{0}{ABC}\ket{000} \hspace*{-1pt}=\hspace*{-1pt}
     2^2\, 3\, f(r)\, g(r)\,,\label{eq:DM matrix elements 3 qubits 000}\\[2mm]
     &\bra{001}\rho\suptiny{0}{0}{\rm G}\subtiny{0}{0}{ABC}\ket{001} \hspace*{-1pt}=\hspace*{-1pt}
     \bra{010}\rho\suptiny{0}{0}{\rm G}\subtiny{0}{0}{ABC}\ket{010}
     \hspace*{-1pt}=\hspace*{-1pt}\bra{100}\rho\suptiny{0}{0}{\rm G}\subtiny{0}{0}{ABC}\ket{100}
\hspace*{-1pt}=\hspace*{-1pt}2^3\, 3\, f(r)\, g(r)\,
\tfrac{[67 + 68 \cosh(2r) + 9 \cosh(4r)] \,\sinh^{\hspace*{-1pt}2}(r)}{249 +
  314 \cosh(2r) + 79 \cosh(4r) + 6 \cosh(6r)}\,,\\[2mm]
  &\bra{011}\rho\suptiny{0}{0}{\rm G}\subtiny{0}{0}{ABC}\ket{011} \hspace*{-1pt}=\hspace*{-1pt}
\bra{101}\rho\suptiny{0}{0}{\rm G}\subtiny{0}{0}{ABC}\ket{101}\hspace*{-1pt}=\hspace*{-1pt}
\bra{110}\rho\suptiny{0}{0}{\rm G}\subtiny{0}{0}{ABC}\ket{110}\\[1mm]
&\ \hspace*{-1pt}=\hspace*{-1pt}\tfrac{f(r)^5\, g(r)^3}{2^3\, 3^3}\,
[20558 \hspace*{-1pt}+\hspace*{-1pt} 38274 \cosh(2 r)\hspace*{-1pt}+\hspace*{-1pt}  24384 \cosh(4 r) \hspace*{-1pt}+\hspace*{-1pt}  8539 \cosh(6 r) \hspace*{-1pt}+\hspace*{-1pt} 
   1458 \cosh(8 r) \hspace*{-1pt}+\hspace*{-1pt}  99 \cosh(10 r)] \,\sinh^{\hspace*{-1pt}2}(r),\nonumber\\[2mm]
   &\bra{111}\rho\suptiny{0}{0}{\rm G}\subtiny{0}{0}{ABC}\ket{111} \hspace*{-1pt}=\hspace*{-1pt}
   \tfrac{f(r)^7\, g(r)^4}{2^5\, 3^5}\,
   [9216316 + 15789701 \cosh(2 r) + 9730682 \cosh(4 r) +
   4155731 \cosh(6 r)\label{eq:DM matrix elements 3 qubits 111}\\[1mm]
   &\hspace*{3.5cm}+ 1182212 \cosh(8 r) + 213057 \cosh(10 r) +
   22086 \cosh(12 r) + 999 \cosh(14 r)] \,\sinh^{\hspace*{-1pt}4}(r),
   \nonumber\\[2mm]
     &\bra{000}\rho\suptiny{0}{0}{\rm G}\subtiny{0}{0}{ABC}\ket{011} \hspace*{-1pt}=\hspace*{-1pt}
     \bra{000}\rho\suptiny{0}{0}{\rm G}\subtiny{0}{0}{ABC}\ket{101} \hspace*{-1pt}=\hspace*{-1pt}
     \bra{000}\rho\suptiny{0}{0}{\rm G}\subtiny{0}{0}{ABC}\ket{110} \hspace*{-1pt}=\hspace*{-1pt}
     \bra{011}\rho\suptiny{0}{0}{\rm G}\subtiny{0}{0}{ABC}\ket{000} \hspace*{-1pt}=\hspace*{-1pt}
     \bra{101}\rho\suptiny{0}{0}{\rm G}\subtiny{0}{0}{ABC}\ket{000} \hspace*{-1pt}=\hspace*{-1pt}
     \bra{110}\rho\suptiny{0}{0}{\rm G}\subtiny{0}{0}{ABC}\ket{000} 
     \nonumber\\[1.5mm]
     &\hspace*{2.47cm}=\,
     f(r)^3\, g(r)^2\,[19 + 16 \cosh(2 r) + \cosh(4 r)]\, \sinh(2 r)\,,\\[2mm]
     &\bra{001}\rho\suptiny{0}{0}{\rm G}\subtiny{0}{0}{ABC}\ket{010} \hspace*{-1pt}=\hspace*{-1pt}
     \bra{001}\rho\suptiny{0}{0}{\rm G}\subtiny{0}{0}{ABC}\ket{100} \hspace*{-1pt}=\hspace*{-1pt}
     \bra{010}\rho\suptiny{0}{0}{\rm G}\subtiny{0}{0}{ABC}\ket{100} \hspace*{-1pt}=\hspace*{-1pt}
     \bra{010}\rho\suptiny{0}{0}{\rm G}\subtiny{0}{0}{ABC}\ket{001} \hspace*{-1pt}=\hspace*{-1pt}
     \bra{100}\rho\suptiny{0}{0}{\rm G}\subtiny{0}{0}{ABC}\ket{010} \hspace*{-1pt}=\hspace*{-1pt}
     \bra{100}\rho\suptiny{0}{0}{\rm G}\subtiny{0}{0}{ABC}\ket{001} \nonumber\\[1.5mm]
     &\hspace*{2.47cm}=\,
     -\,f(r)^3\, g(r)^2\,2\,[2 + \cosh(2 r)]\, \sinh^{\hspace*{-1pt}2}(2r)\,, \\[2mm]
     &\bra{001}\rho\suptiny{0}{0}{\rm G}\subtiny{0}{0}{ABC}\ket{111} \hspace*{-1pt}=\hspace*{-1pt}
     \bra{010}\rho\suptiny{0}{0}{\rm G}\subtiny{0}{0}{ABC}\ket{111} \hspace*{-1pt}=\hspace*{-1pt}
     \bra{100}\rho\suptiny{0}{0}{\rm G}\subtiny{0}{0}{ABC}\ket{111} \hspace*{-1pt}=\hspace*{-1pt}
     \bra{111}\rho\suptiny{0}{0}{\rm G}\subtiny{0}{0}{ABC}\ket{001} \hspace*{-1pt}=\hspace*{-1pt}
     \bra{111}\rho\suptiny{0}{0}{\rm G}\subtiny{0}{0}{ABC}\ket{010} \hspace*{-1pt}=\hspace*{-1pt}
     \bra{111}\rho\suptiny{0}{0}{\rm G}\subtiny{0}{0}{ABC}\ket{100} \nonumber\\[1.5mm]
     &\hspace*{2.47cm}=\,
     \tfrac{f(r)^5\, g(r)^2}{2}\,[54 \cosh(r) + 17 \cosh(3 r) + \cosh(5 r)]\, \sinh^{\hspace*{-1pt}3}(r)\,,\\[2mm]
     &\bra{011}\rho\suptiny{0}{0}{\rm G}\subtiny{0}{0}{ABC}\ket{101} \hspace*{-1pt}=\hspace*{-1pt}
     \bra{011}\rho\suptiny{0}{0}{\rm G}\subtiny{0}{0}{ABC}\ket{110} \hspace*{-1pt}=\hspace*{-1pt}
     \bra{101}\rho\suptiny{0}{0}{\rm G}\subtiny{0}{0}{ABC}\ket{011} \hspace*{-1pt}=\hspace*{-1pt}
     \bra{101}\rho\suptiny{0}{0}{\rm G}\subtiny{0}{0}{ABC}\ket{110} \hspace*{-1pt}=\hspace*{-1pt}
     \bra{110}\rho\suptiny{0}{0}{\rm G}\subtiny{0}{0}{ABC}\ket{101} \hspace*{-1pt}=\hspace*{-1pt}
     \bra{110}\rho\suptiny{0}{0}{\rm G}\subtiny{0}{0}{ABC}\ket{011} \nonumber\\[1.5mm]
     &\hspace*{2.47cm}=\,
     \tfrac{f(r)^5\, g(r)^2}{4}\,[33+ 22\cosh(2r) - \cosh(4 r)]\, \sinh^{\hspace*{-1pt}2}(2r)\,, \label{eq:DM matrix elements 3 qubits 011,101}
\end{align}
\end{subequations}
\end{small}
while all other (off-diagonal) density-matrix elements vanish in the subspace with at most one excitation in each mode. 
Inserting these values into the witness Ineq.~(\ref{ineq FS state SM}) and numerically evaluating it, we find that the inequality is violated for all values of $r$ in the range $0<r<r_{0}'$ with $r_{0}^{\prime}\approx0.284839$.
\subsection{Entanglement in the three-qubit subspace}\label{SM:3.4}

Using the density-matrix elements in Eqs.~(\ref{eq:DM matrix elements 3 qubits 000})-(\ref{eq:DM matrix elements 3 qubits 011,101}) of the three-mode state $\rho\suptiny{0}{0}{\rm G}\subtiny{0}{0}{ABC}$ we can further project the state into the subspace spanned by the Fock states with at most one excitation in each mode. 
This procedure results in a three-qubit state $\rho\suptiny{0}{0}{\rm QB}\subtiny{0}{0}{ABC}=\Lambda[\rho\suptiny{0}{0}{\rm G}\subtiny{0}{0}{ABC}]$ whose density-matrix elements are obtained by dividing all matrix elements in Eqs.~(\ref{eq:DM matrix elements 3 qubits}) by the sum of the eight diagonal elements in Eqs.~(\ref{eq:DM matrix elements 3 qubits 000})-(\ref{eq:DM matrix elements 3 qubits 111}),
\begin{align}
    \bra{ijk}\rho\suptiny{0}{0}{\rm QB}\subtiny{0}{0}{ABC}\ket{i'j'k'} &=
    \frac{\bra{ijk}\rho\suptiny{0}{0}{\rm G}\subtiny{0}{0}{ABC}\ket{i'j'k'}}{\sum\limits_{\substack{l,m,n \\ =0,1}}\!
    \bra{lmn}\rho\suptiny{0}{0}{\rm G}\subtiny{0}{0}{ABC}\ket{lmn}}\,.
\end{align}
Such a local filtering $\Lambda$ can increase the entanglement of the state, but it cannot create (genuine multipartite) entanglement for any state that is (bi)separable to begin with.

With this in mind, we can check the PPT criterion~\cite{Peres1996,HorodeckiMPR1996} for this state. We find that the operator obtained by transposing any single qubit has a negative eigenvalue when $0<r<r_1$ with $r_1$ as in Eq.~(\ref{eq:r1}). 
From the symmetry of the state and the fact that $\Lambda$ cannot create entanglement, we can thus infer that $\rho\suptiny{0}{0}{\rm QB}\subtiny{0}{0}{ABC}$, and hence $\rho\suptiny{0}{0}{\rm G}\subtiny{0}{0}{ABC}$ must be fully inseparable for $0<r<r_{1}$.
It then follows from Ref.~\cite{PalazuelosDeVicente2022} that there is some $k\geq2$ such that $(\rho\suptiny{0}{0}{\rm QB}\subtiny{0}{0}{ABC})^{\otimes k}$ is GME (if ${\rho\suptiny{0}{0}{\rm QB}\subtiny{0}{0}{ABC}}$ is not already GME) in the same region, and since $(\rho\suptiny{0}{0}{\rm QB}\subtiny{0}{0}{ABC})^{\otimes k}=\Lambda^{\otimes k}[(\rho\suptiny{0}{0}{\rm G}\subtiny{0}{0}{ABC})^{\otimes k}]$, also ${\rho\suptiny{0}{0}{\rm G}\subtiny{0}{0}{ABC}}^{\otimes k}$ must be at least GME activatable for all values of $r$ between $r_0$ and $r_1$, but we do not (yet) know if it is GME on the single-copy level in this parameter range.\\[-2.5mm]

The three-qubit state $\rho\suptiny{0}{0}{\rm QB}\subtiny{0}{0}{ABC}$ gives us more opportunities to detect GME in the Gaussian state $\rho\suptiny{0}{0}{\rm G}\subtiny{0}{0}{ABC}$. 
A straightforward method to use is an entanglement witness known as a fully decomposable witness $W$~\cite{JungnitschMoroderGuehne2011a}, which generalizes the PPT criterion for the detection of GME.
For every subset $M$ of parties, we can define an operator
\begin{align}\label{eq:fdecwit}
    W = P_M + Q_M^{T_M},
\end{align}
where $P_M$ and $Q_M$ are positive semi-definite operators and $T_M$ signifies partial transposition with respect to the subsystem $M$.\\[-2.5mm]

The fully decomposable witness is non-negative on all states that are convex combinations of states with positive partial transposition for all possible bipartitions. 
The set of these states contains all biseparable states and some GME states because the PPT criterion is not sufficient for separability in dimensions of the joint Hilbert space higher than $2\times 3$~\cite{Horodecki1997}. An advantage of fully decomposable witnesses is the possibility of evaluating them using the convex optimization technique of semi-definite programming, which allows us to optimize the result over the whole set of fully decomposable witnesses.\\[-2.5mm] 

Applying this technique for our three-qubit state using publicly available Python code~\cite{ProvaznikGithub}, we detect GME in the three-qubit state for all values of $r$ in the range $0<r<r_0$ with $r_0=0.575584$. 
This result indicates that the original Gaussian state is GME at least in the range $0<r<r_0$.


\section{Projection of CV FIB states to finite dimensions}\label{SM:4}

In this section, we will argue that results in~\cite{SperlingVogel2009} guaranteeing points~\ref{first} and~\ref{second} can be extended so that for every CV FIB state one can always find a local projection such that it projects it to a FIB state in finite dimension, as stated in Corollary 1 in Sec.~\ref{sc:GMEactivationinfinitedimension}.\\

Following the proof of Theorem 2 in Ref.~\cite{SperlingVogel2009}, we first note that each projector $P_{d}=\bigotimes_{j=1}^{N}P_{d}^{j}$ belongs to a sequence of local projectors indexed by the dimension $d$ of the subspace that they project into, which converges pointwise to the identity operator on the original infinite-dimensional Hilbert space as $d\rightarrow\infty$. 
In particular, this convergence of the projectors to the identity ensures that every (FIB) state $\rho$ satisfies
\begin{align}\label{fixedsequence}
        \rho\,=\,\lim_{d \to \infty} P_{d}\,\rho\,P_{d}.
\end{align}
Since $\rho$ is FIB, it must be entangled with respect to any bipartition $M(i)$, and so for any $M(i)$, there must exist an entanglement witness, i.e., a bounded Hermitian operator $W_{i}$ such that
\begin{align}
    \tr(\rho\,W_{i})    &<\,0\,,
\end{align}
while for all states $\sigma_{i}$ that are separable with respect to the bipartition $M(i)$ we have
\begin{align}
    \tr(\sigma_{i}\,W_{i})    &\geq\,0\,.
    \label{eq:witness sep state}
\end{align}
If we then assume that the finite-dimensional projections $\rho_{d}$ for all $d$ are separable with respect to any of the bipartitions $M(i)$ and that $\tr(\rho\,W_{i})$ are all continuous linear functions of $\rho$, then for all entanglement witnessses $W_{i}$ we have
\begin{align}
    \tr(\rho\,W_{i}) &=\,
    \tr(\lim_{d \to \infty} P_{d}\,\rho\,P_{d}\,W_{i})=\,
    \lim_{d \to \infty}\,\underbrace{\tr( P_{d}\,\rho\,P_{d}\,W_{i})}_{\geq\,0}\,\geq\,0\,,
\end{align} 
and hence a contradiction with the premise that $\rho$ is entangled with respect to all bipartitions. Consequently, for every bipartition $M(i)$ there must be a finite dimension $d_{i}$ for which $\tr(\rho_{d_{i}}\,W_{i})<0$, i.e., for which $\rho_{d_{i}}$ is entangled with respect to $M(i)$. Since all of the states are within the same sequence and the entanglement of 
$\rho_{d_{i}}$ implies entanglement of $\rho_{d_{i}'}$ with $d_{i}'\geq d_{i}$ by point 2, there must be some finite dimension $d_{\mathrm{max}}=\max_{i}d_{i}$ such that $\rho_{d_{\mathrm{max}}}$ is fully inseparable. 
Since the projections are all local and $\rho$ was biseparable to begin with, also $\rho_{d_{\mathrm{max}}}$ must be biseparable, and thus FIB, which concludes the proof of Corollary 1 in Sec.~\ref{sc:GMEactivationinfinitedimension}.\hfill

\end{document}